\documentclass[12pt]{article}
\usepackage[body={18cm, 23cm, centeblack}]{geometry}
\usepackage{amsmath,amssymb,amsthm,amscd,cite,comment,amsfonts,indentfirst,color,setspace,bbold}

\usepackage[debug,pageanchor=false]{hyperref}
\hypersetup{colorlinks=true,linktocpage,breaklinks,
	urlcolor=black,
	linkcolor=black,
	citecolor=black
}

\usepackage{tikz}
\usetikzlibrary{arrows}
\usetikzlibrary{shapes.misc}

\tikzset{cross/.style={cross out, draw=black, minimum size=2*(#1-\pgflinewidth), inner sep=0pt, outer sep=0pt},
	cross/.default={5pt}}
\usepackage[ normalem]{ulem}
\usepackage{tocloft}

\usepackage{xcolor}

\numberwithin{equation}{section}

\selectfont

\def\a{\alpha} 
\def\b{\beta} 
 
\def\d{\delta} 
\def\e{\epsilon}

\def\h{\eta}

\def\l{\lambda} 
\def\m{\mu}
\def\n{\nu} 
 
\def\r{\rho}
\def\q{\theta}

\def\f{\phi}

\def\Q{\Theta}

\def\L{\Lambda}

\def\W{\Omega}

\def\ba{\bar{a}}

\def\bmu{\bar{\mu}}

\def\bi{\bar{i}}
\def\bj{\bar{j}}
\def\bg{\bar{g}}
\def\bm{\bar{m}}
\def\bn{\bar{n}}

\def\brho{\bar{\rho}}

\def\o{\overline}

\def\fr{\frac}  \def\dt{\partial}

\def\mc{\mathcal}

\def\mE{\mathcal{E}}
\def\mF{\mathcal{F}}

\def\mL{\mathcal{L}}
\def\mM{\mathcal{M}}

\def\SS{\mathbb{S}}

\def\XX{\mathbb{X}}
\def\RR{\mathbb{R}}

\def\OO{\mathrm{O}}

\begin{document}
\renewcommand{\refname}{\begin{center}References\end{center}}
	
\begin{titlepage}
		
	\vfill
	\begin{flushright}

	\end{flushright}
		
	\vfill
	
	\begin{center}
		\baselineskip=16pt
		{\Large \bf 
		 Polyvector deformations in \\eleven-dimensional supergravity 
		}
		\vskip 1cm
		    Kirill Gubarev$^{a,b}$\footnote{\tt kirill.gubarev@phystech.edu },
		    Edvard T. Musaev$^{a,c}$\footnote{\tt musaev.et@phystech.edu},
		\vskip .3cm
		\begin{small}
			{\it 
			    $^a$Moscow Institute of Physics and Technology, \\
				Institutskii per. 9, Dolgoprudny, 141700, Russia,\\
				$^b$Institute for Theoretical and Experimental Physics,\\ B. Cheremushkinskaya, 25, 117218, Moscow, Russia \\
				$^c$Kazan Federal University, Institute of Physics, \\Kremlevskaya 16a, Kazan, 420111, Russia\\
			}
		\end{small}
	\end{center}
		
	\vfill 
	\begin{center} 
		\textbf{Abstract}
	\end{center} 
	\begin{quote}
         We consider 3- and 6-vector deformations of 11-dimensional supergravity backgrounds of the form $M_5\times M_6$ admitting at least 3 Killing vectors. Using flux formulation of the E${}_{6(6)}$ exceptional field theory we derive (sufficient) conditions for the deformations to generate a solution. In the group manifold case these generalisations of the classical Yang-Baxter equation for the case of r-matrices with 3 and 6 indices are shown to reproduce  those obtained from exceptional Drinfeld algebra for E${}_{6(6)}$. In general we see an additional constraint, which might be related to higher exceptional Drinfeld algebras.
	\end{quote} 
	\vfill
	\setcounter{footnote}{0}
\end{titlepage}
	
\clearpage
\setcounter{page}{1}

\tableofcontents


\section{Introduction}

Vacua of string theory understood as a perturbative formulation of the non-linear two-dimensional sigma-model are known to be represented by a vast landscape of 10-dimensional manifolds equipped by various gauge fields: Kalb-Ramond 2-form and Ramond-Ramond $p$-form gauge fields. In the full non-perturbative formulation one finds that the space of string vacua is mostly populated by 11-dimensional manifolds and 10-dimensional backgrounds represent points with small string coupling constant (see \cite{Townsend:1996xj} for more detailed review). The set of string vacua possesses huge amount of various symmetries that prove useful in  better understanding of its structure. In particular one finds T-duality symmetries: abelian \cite{Buscher:1987sk, Buscher:1987qj}, non-abelian \cite{delaOssa:1992vc} and more generally Poisson-Lie dualities \cite{Klimcik:1995dy, Klimcik:1995ux,Klimcik:1995jn}, which relate backgrounds that are indistinguishable from the perturbative string point of view. Non-perturbatively symmetries get enhanced to (abelian) U-dualities, that can be understood as transformations  relating toroidal backgrounds equivalent from the point of view of the membrane \cite{Hull:1994ys,Witten:1995zh}. Certain progress towards defining non-abelian generalisation of U-dualities have been made recently in \cite{Sakatani:2019zrs,Malek:2019xrf,Sakatani:2020iad,Musaev:2020bwm,Sakatani:2020wah}.

At the level of low-energy theory of background fields such duality symmetries appear as solution generating transformations. More generally one is interested in transformations that keep the string sigma-model in a consistent vacuum, however changing it in a controllable way. Particularly interesting examples are based on manifolds with an AdS factor, which are known to be holographically dual to superconformal field theories. While in general CFTs are isolated point in the space of couplings corresponding to fixed point of renormalisation group flow, SCFTs belong to a family of theories connected by varying couplings. Adding exactly marginal operators to a theory will preserve conformal symmetry at the quantum level and move the corresponding point in the space of couplings along the so-called conformal manifold \cite{Cordova:2016xhm}. Certain progress in understanding of the structure of conformal manifold can be made by investigating the gravitational side of the AdS/CFT correspondence. Indeed, given a set of exactly marginal operators that deform a SCFT keeping it on a conformal manifold, there exists  a family of dual AdS solutions related by deformations of metric, dilaton and $p$-form gauge fields. The well-known example is provided by $\b$-deformations of $D=4$ $\mc{N}=4$ SYM whose gravity dual is a bi-vector abelian (TsT) deformation along two of three U(1) isometry directions of $\SS^5$ \cite{Lunin:2005jy}. In the similar fashion considering AdS${}_4\times \SS^7$ one is able to pick three U(1) directions of the seven-sphere to construct a trivector deformation of ABJM theory. For a general formula for TsT transformations of gauge theories see \cite{Imeroni:2008cr}.

Generalising the results known for TsT deformations one naturally gets interested in bi- and tri-vector deformations along a set of non-commuting Killing vectors. As the most symmetric example here one finds deformations of two-dimensional sigma models preserving integrability, e.g. $\h$-deformation of the Green-Schwarz superstring on AdS${}_5\times \SS^5$ constructed in  \cite{Arutyunov:2013ega, Hoare:2018ngg}. Depending on the choice of Dynkin diagram for the corresponding superalgebra this gives rise to the so-called ABF background \cite{Arutyunov:2015qva}, which solves equations of motion of generalised supergravity \cite{Arutyunov:2015mqj,Hoare:2015wia,Wulff:2016tju}, or to normal supergravity backgrounds. To depart from backgrounds given by group manifolds and coset spaces one generalises the procedure and for a general deformation parametrised by a bi-vector $\Q=r^{ij}k_i \wedge k_j$ obtains
\begin{equation}
\label{eq:openclosed}
    (g+b)^{-1}=(G+B)^{-1}+\Q,
\end{equation}
where $G,B$ and $g,b$ are the metrices and the 2-form fields for the initial and deformed backgrounds  respectively \cite{Araujo:2017jkb,Araujo:2017jap,Borsato:2018idb}. Although at this level the deformation is consideblack as a transformation of (generalised) supergravity solutions without direct reference to two-dimensional sigma-models, both initial and deformed field configurations could be understood as consistent sigma-model backgrounds. Both the sigma-model (for coset spaces) and field theory (for general manifolds with Killing vectors) approaches show that a deformed background is a solution of (generalised) supergravity equations when the r-matrix  $r^{ij}$ satisfies classical Yang-Baxter equation (CYBE) \cite{Bakhmatov:2017joy,Bakhmatov:2018bvp}
\begin{equation}
    r^{k [i_1 }r^{i_2 |l|}f_{kl}{}^{i_3]} = 0.
\end{equation}
Here $f_{ij}{}^k$ are structure constants of the algebra of Killing vectors.

It is important to notice, that the transformation \eqref{eq:openclosed} appears essentially non-linear only due to the bad choice of variables, and becomes a linear O$(10,10)$ transformation when written in terms of generalised metric, i.e. a representative of the coset $\OO(10,10)/\OO(1,9)\times \OO(9,1)$. For this reason deriving classical Yang-Baxter equation from the standard supergravity formalism faces huge technical difficulties and has been done in \cite{Bakhmatov:2018apn,Bakhmatov:2017joy} only in the second order in $\Q$. More natural appears the formalism of Double Field Theory \cite{Hohm:2010jy,Hohm:2010pp}, where the generalised metric is the canonical variable and which allowed full proof in \cite{Bakhmatov:2018bvp} that CYBE is sufficient to end up with  a solution. Moving to tri-vector deformations one naturally employs the formalism of exceptional field theory (ExFT) for precisely the same reasons: the generalised metric transforms linearly under deformations. The generalisation of the deformation map \eqref{eq:openclosed} obtained in the formalism of SL(5) ExFT in \cite{Bakhmatov:2019dow} finds the same interpretation as an open-closed membrane map \cite{Berman:2001rka}. Examples of non-abelian tri-Killing deformations based on the open-closed membrane map, or equivalently a specially defined SL(5) transformation, have been provided for the AdS${}_4\times \SS^7$ background in \cite{Bakhmatov:2020kul}.

One may naturally ask, whether the condition that a tri-vector deformation generates a solution of supergravity is equivalent to some algebraic condition generalising CYBE. One first notices that CYBE appears when deforming generators of a Manin triple $(T_i, T^i, \h)$ representing a Drinfeld double algebra by r-matrix
\begin{equation}
    \begin{aligned}
       T_i & \to T_i,\\
       T^i & \to T^i + r^{ij}T_j,
    \end{aligned}
\end{equation}
and requiring the deformed generators to also form a Drinfeld double algebra. Similarly deforming generators of an exceptional Drinfeld algebra (EDA) by tri- and six-vector tensors $\r^{i_1\dots i_3}$, $\r^{i_1\dots i_6}$ and restricting the deformed generators to form an EDA, one arrives at a set of conditions on $\r$-tensors. Exceptional Drinfeld algebra based on the SL(5) group has been constructed in \cite{Sakatani:2019zrs, Malek:2019xrf} and for the group E${}_{6(6)}$ this has been done in \cite{Malek:2020hpo}. Conditions on the 3-vector deformation tensor $\r^{i_1i_2i_3}$ derived for the SL(5) EDA in \cite{Sakatani:2019zrs} are  equivalent to the unimodularity condition
\begin{equation}
    \r^{kl [i_1}f_{kl}{}^{i_2]}=0,
\end{equation}
that is due to the dimension $d=4$ of the internal manifold, which appear to be too small to embed 3-vector deformations. The situation is the same as for bi-vector deformations in dimension $d=3$ where one gets only the unimodularity condition.

In contrast, deformations consideblack inside the E${}_{6(6)}$ EDA are subject to a non-trivial constraint, which is supposed to generalise classical Yang-Baxter deformation. In this work we investigate whether this condition is sufficient for a deformation to be a solution generating transformation as for the bi-vector case. For this we start with providing short review of the E${}_{6(6)}$ extended geometry in Section \ref{sec:truncation}. In Section \ref{sec:defsflux} we define deformation map for E${}_{6(6)}$ generalised vielbein and investigate transformation of generalised fluxes under the map. We find, the generalised Yang-Baxter equation of \cite{Malek:2020hpo} as a sufficient condition for the fluxes to stay undeformed, as well as an additional constraint.

\section{Truncation of 11D supergravity}
\label{sec:truncation}

Bi-vector deformation of backgrounds of non-linear sigma-model is given by the non-linear map \eqref{eq:openclosed}, whose form cannot be called self-evident. Due to this non-linear nature of the map explicit check that this is a solution generating transformation is a highly non-trivial task, as it has been demonstrated in \cite{Bakhmatov:2017joy}. Choosing correct representation of degrees of freedom allows to turn deformation map into a linear transformation. Hence, for bi-vector deformations one organises NS-NS fields into the generalised metric of double field theory parametrising the coset $O(d,d)/O(d)\times O(d)$ and the invariant dilaton, and R-R fields into an O$(d,d)$ spinor. In this case the transformation becomes simply an O$(d,d)$ rotation and showing its solution-generating nature becomes a straightforward task (see \cite{Sakamoto:2017cpu,Borsato:2018idb,Catal-Ozer:2019tmm} for more details).

Similarly, as it has been shown in  \cite{Bakhmatov:2019dow} tri-vector deformations of 11-dimensional background can be defined as an SL(5) transformation of generalised  metric of the corresponding exceptional field theory. Below we show that for 3-vector deformations within the SL(5) theory to generate a solution unimodularity is sufficient. Hence, no generalised Yang-Baxter equation is generated both in the field theory and in the SL(5) EDA. Here we consider the E${}_{6(6)}$ exceptional field theory that allows 3- and 6-vector deformations and, according to the analysis of exceptional Drinfeld algebra of \cite{Malek:2020hpo} a non-trivial generalisation of the classical Yang-Baxter equation exists.

Detailed presentation of the E${}_{6(6)}$ exceptional field theory can be found in \cite{Hohm:2013vpa,Musaev:2014lna,Baguet:2015xha} that includes bosonic and fermionic field content, supersymmetry transformations, full Lagrangian and truncations to the 11-dimensional and Type IIB supergravities. For review of the recent progress see \cite{Berman:2020tqn,Musaev:2019zcr}. For our purposes here  we stress the following defining features of the theory:
\begin{itemize}
    \item ExFT is an E$_{d(d)}$-covariant background independent theory combining \emph{full} 11-dimensional and Type IIB supergravities (no blackuction);
    \item field content is represented by tensors of GL(11-d) taking values in certain representations of the U-duality group;
    \item section condition, restricting dependence of fields on the total $5+27$ coordinates is requiblack. In what follows we assume the standard solution of the section constraint leaving only dependence on $5+6$ coordinates. 
\end{itemize}
Hence, for the purposes of this work exceptional field theory simply provides a convenient rewriting of degrees of freedom of 11-dimensional supergravity, turning polyvector deformations into a linear map.

Construction of fields and the Lagrangian of the E${}_{6(6)}$ exceptional field theory starting from the 5+6 split of the 11-dimensional supergravity is given in details in \cite{Hohm:2013vpa}. To introduce notations and for further reference we provide a brief overview of the setup. Bosonic field content of 11-dimensional supergravity  consists of the elfbein $\hat{E}_{\hat{\m}}{}^{\hat{\a}}$ and the 3-form potential $C_{\hat{\mu}_1\hat{\mu}_{2}\hat{\mu}_{3}}$. Keeping full dependence on all of the 11 coordinates $x^{\hat{\m}}$ one splits the fields into tensors in 5-dimensions and organises them into multiplets of E${}_{6(6)}$. For the latter one has to follow the dualisation prescription of \cite{Cremmer:1997ct}. Decomposing 11-dimensional indices as $\hat{\m}=(\m,m)$, $\hat{\alpha}=(\o{\m},a)$ one  parametrises the elfbein in the following upper-triangular form
\begin{equation}
\label{KKgauge}
  \hat{E}_{\hat{\mu}}{}^{\hat{\alpha}} \ = \ 
  \left(\begin{array}{cc} e^{-\frac13}g_{\mu}{}^{\o{\m}} &
  A_{\mu}{}^{m} e_{m}{}^{a} \\ 0 & e_{m}{}^{a}
  \end{array}\right)\,, 
\end{equation}
and blackefines fields arising from the 3-form potential as
  \begin{equation}
   \begin{split}
    A_{mnk} \ &= \ C_{mnk}\;, \\
    A_{\mu\,mn} \ &= \ C_{\mu mn}-A_{\mu}{}^k\,C_{kmn}\;, \\
    A_{\mu\nu\,m} \ &= \ C_{\mu\nu m}-2 A_{[\mu}{}^n\,C_{\nu]mn}+A_{\mu}{}^n A_{\nu}{}^k\,C_{mnk}\;, \\
    A_{\mu\nu\rho} \ &= \ C_{\mu\nu\rho}-3 A_{[\mu}{}^m\,C_{\nu\rho]m}+3 A_{[\mu}{}^m A_{\nu}{}^n\, C_{\rho]mn}
    -A_{\mu}{}^m A_{\nu}{}^n A_{\rho}{}^k\,C_{mnk}\;.
   \end{split}
   \label{comp3form}
  \end{equation}

The resulting fields
\begin{equation}\label{s5+6fields}
    \begin{aligned}
     \{ & g_{\mu}{}^{\o{\m}}, && A_{\mu}{}^{m}, && e_{m}{}^{a}, && A_{mnk}, && A_{\m nk}, && A_{\m \n k}, && A_{\m \n \r} \}
    \end{aligned}
\end{equation}
transform appropriately under slpitted 11-dimensional diffeomorphisms $\xi^{\hat{\m}}=(\xi^{\m},\L^{m})$. To organise these into multiplets of E${}_{6(6)}$ one has to dualise all forms to the lowest possible rank. Hence, the two forms $A_{\m \n m}$ get dualised intro 1-forms and can be collected with $A_\m{}^m$ and $A_{\m mn}$ into the vector $A_\m{}^M$ of ExFT transforming in the $\bf 27$ of E${}_{6(6)}$; the 3-form $A_{\m\n\r}$ after dualization contributes to the scalar coset. The 6-dimensional space gets extended to include coordinates corresponding to winding modes of the M2- and M5-branes and is now parametrised by $\XX^M$ transforming as the $\bf 27$ under global U-duality transformations. Local coordinate transformations on such extended space are given by the so-called generalised Lie derivative
\begin{equation}\label{LieV}
    \mL_{\L} V^{M} = \L^K \dt_K V^{M} - V^{K} \dt_K \L^M   + 10 d^{MKR} d_{NLR} V^{N} \dt_K \L^L- \bigg(\l -\frac13\bigg) V^{M} \dt_K \L^K,
\end{equation}
where $V^M$ denote components of some generalised vector of weight $\l$ and $d^{MNK}$  and $d_{MNK}$ are the invariant tensors of E${}_{6(6)}$. For such defined transformations to form a closed algebra one imposes the section constraint
\begin{equation}\label{sectioncondition}
    d^{M N K} \dt_N \bullet \dt_M \bullet = 0,
\end{equation}
where bullets denote any fields and their combinations. The above condition has two (maximal) inequivalent solutions corresponding to embeddings of the 11-dimensional and Type IIB 10-dimensional supergravity. We will be working with the former, i.e. always take into account decomposition of $\mathfrak{e}{}_{6(6)}$ irreps under its subalgebra $\mathfrak{gl}(6)$. For the coordinates $\XX^M$ this reads
\begin{equation}\label{e6reprbreakgencoord}
    \mathbb{X}^{M} = (x^{m},x_{mn},x^{\bm}),
\end{equation}
where $m,n = \o{1,6}$, $\bm,\bn = \o{1,6}$. We refer to appendix \ref{NotesRefs} for the used index notations and conventions and notice here, that barblack indices label the same $\mathbf{6}$ of E${}_{6(6)}$ and these are distinguished from unbarblack small Latin indices for technical convenience. In what follows we always assume 
\begin{equation}\label{sectioncondition1}
    \dt_{\bm}f=0,\, \dt^{mn}f=0,
\end{equation}
for any field $f$ of the theory. Upon such decomposition  non-vanishing components of the symmetric invariant tensor can be written as
\begin{equation}\label{Dtensors}
    \begin{aligned}
       d^{\bn n}{}_{m_1 m_2} & = \frac{1}{\sqrt{5}} \delta_{[m_1}{}^{\bn} \delta_{m_2]}{}^{n}, &&
       d_{n_1 n_2 n_3 n_4 n_5 n_6} &&& = \frac{1}{4\sqrt{5}} \epsilon_{n_1 n_2 n_3 n_4 n_5 n_6},\\
       d_{\bn n}{}^{m_1 m_2} & = \frac{1}{\sqrt{5}} \delta^{[m_1}{}_{\bn} \delta^{m_2]}{}_{n}, &&
       d^{n_1 n_2 n_3 n_4 n_5 n_6} &&& = \frac{1}{4\sqrt{5}} \epsilon^{n_1 n_2 n_3 n_4 n_5 n_6}.
    \end{aligned}
\end{equation}
Finally, the (bosonic) E$_{6(6)}$ ExFT field content reads
\begin{equation}\label{E6fields}
    \begin{aligned}
     \{ & g_{\m\n}, && \mM_{MN}, && A_\m\,^{M}, && B_{\m\n M}\},
    \end{aligned}
\end{equation}
where $g_{\m\n}$ is  the metric of the external space, $\mM_{MN}$ is the so-called generalized metric parametrising the scalar coset, $A_\m\,^{M}$ is a generalized connection and $B_{\m\n M}$ is a set of two-forms.

It is convenient to turn from the generalized metric $\mM_{MN}$ to generalized vielbeins $\mE_{M}^{A} $ defined as
\begin{equation}\label{MthroughE}
    \mM_{M N} = \mE_{M}{}^{A} \mE_{N}{}^{B} \mM_{A B},
\end{equation}
where $\mM_{A B}$ is a constant matrix (unity, for concreteness). In terms of fields of 11-dimensional supergravity  the generalized vielbein $\mE_{M}^{A}$ can be parametrized as follows
\begin{equation}\label{uptrE6}
    \mE_{M}{}^{A} = e^{\frac13}
    \begin{bmatrix}
        e_{m}^{a} & -\frac{1}{\sqrt{2}}e_{m}^{b} C_{b a_1 a_2} & \frac{1}{2} e_{m}^{\ba} U + \frac{1}{4} e_{n}^{\ba} C_{m b_1 b_2} V^{n b_1 b_2}\\
        
        0  & e_{[a_1}^{m_1} e_{a_2]}^{m_2} & - \frac{1}{\sqrt{2}} e_{b_1}^{m_1} e_{b_2}^{m_2} V^{b_1 b_2 \ba} \\
        
        0 & 0 & e^{-1} e_{\bm}^{\ba} 
    \end{bmatrix},
\end{equation}
where $e = \det(e_{m}^{a})$ and $V^{m_1 m_2 m_3} = \frac{e^{-1}}{3!} \epsilon^{m_1 m_2 m_3 n_1 n_2 n_3} C_{n_1 n_2 n_3}$. The  scalar degree of freedom $U = \frac{e^{-1}}{6!} \epsilon^{n_1 ... n_6} C_{n_1 ... n_6}$ comes from dualization of the three-form $A_{\m\n\r}$ and the procedure directly relating these two can be found in \cite{Hohm:2013vpa}. For the inverse vielbein one has
\begin{equation}\label{invuptrE6}
    \mE^{M}{}_{A} = e^{-\frac13}
    \begin{bmatrix}
        e^{m}_{a} & 0 & 0\\
        
        \frac{1}{\sqrt{2}}e^{n}_{a} C_{n m_1 m_2}  & e^{[a_1}_{m_1} e^{a_2]}_{m_2} & 0 \\
        
        - \frac{e}{2} e^{\bm}_{a} U + \frac{e}{4} e_{a}^{l} C_{l k n} V^{\bm k n} & \frac{e}{\sqrt{2}} V^{a_1 a_2 \bmu} & e e^{\bm}_{\ba} 
    \end{bmatrix}.
\end{equation}
Such defined generalised vielbein is a generalised vector of weight $\l=0$, that is necessary for the full Lagrangian of ExFT to be invariant
\begin{equation}\label{LieE}
    \mL_{\L} \mE^{M}{}_{A} = \L^K \dt_K \mE^{M}{}_{A} - \mE^{K}{}_{A} \dt_K \L^M + 10 d^{MKR} d_{NLR} \mE^{N}{}_{A} \dt_K \L^L + \fr13  \mE^{M}{}_{A} \dt_K \L^K .
\end{equation}

In this work we will be focusing only at the scalar sector of the theory, i.e. only at the fields entering the generalised vielbein. As in the SL(5) theory \cite{Bakhmatov:2020kul} consistent truncation requires to also keep track of determinant of the external metric $g_{\m\n}$. Below we discuss a rescaling of the generalised metric, that combines all these degrees of freedom and  decouples them from the rest of the fields.

\subsection{Generalised flux formulation}

For  simplification of further discussion we consider only such backgrounds, that can be presented as $M_{11} = M_{6} \times M_{5}$ , where the internal metric $g_{mn}$, the 3-form $C_{m_1 m_2 m_3}$ and the 6-form $C_{m_1 m_2 m_3 m_4 m_5 m_6}$ don't depend on the external coordinates $y^{\m}$. Also we take $A_{\mu}{}^{M}=0$ and $B_{\mu \nu M}=0$. Hence, we consider the following ansatz
\begin{equation}
\label{anzatse}
    \begin{aligned}
       &g_{\m\n}=g_{\m\n}(y^\m, x^m), && \mM_{MN}=\mM_{MN}(x^m), \\
       &A_{\m}{}^{MN}=0, && B_{\m\n\, M} =0.
    \end{aligned}
\end{equation}
Since the external metric is a scalar of non-zero weight under generalised Lie derivative and, as we discuss later, the deformations are given by E${}_{6(6)}$ transformations, the external metric transforms by a rescaling. It is convenient to explicitly factor out the part of non-zero weight $e^{2\f}$ of the external metric and further restrict coordinate dependence as  $g_{\m\n}(y^\m,x^m)=e^{-2\f(x^m)}e^{\frac{2}{9-d}}\bar{g}_{\m\n}(y^\m)$ ($d=6$ for E$_{6(6)}$). The factor  $e^{-2\f(x^m)}$ is possible to combine with the generalised metric of ExFT $\mc{M}_{MN}$ to define
\begin{equation}
\label{rescalings}
    \begin{aligned}
     g_{\m\n}(y^\m,x^m) & = e^{-2\f(x^m)}e^{\frac{2}{9-d}}\bar{g}_{\m\n}(y^\m), && \mM_{MN}  = e^{-2\f}e^{\frac{2}{9-d}} M_{MN}.
     \end{aligned}
\end{equation}
Applied to the SL(5) theory one has $d=4$ and the metric $M_{MN}$ will me precisely that of the truncated theory of \cite{Blair:2014zba}. Such  rescaled generalized metrics  $M_{MN} \in E_{6(6)} \times \mathbb{R}^{+}$ and can be represented in terms of the generalised vielbein $E_M{}^A \in E_{6(6)} \times \mathbb{R}^{+}$ as usual
\begin{equation}\label{metricvielbein}
    M_{MN} = E_M{}^A E_N{}^B M_{AB}.
\end{equation}
In components the generalised vielbein and its inverse read
\begin{equation}\label{uptrE6R}
\begin{aligned}
    E_{M}{}^{A} &= e^{\phi}
    \begin{bmatrix}
        e_{\mu}^{i} & -\frac{1}{\sqrt{2}}e_{\mu}^{k} C_{k i_1 i_2} & \frac{1}{2} e_{\mu}^{\bi} U + \frac{1}{4} e_{\nu}^{\bi} C_{\mu k_1 k_2} V^{\nu k_1 k_2}\\
        
        0  & e_{[i_1}^{\mu_1} e_{i_2]}^{\mu_2} & - \frac{1}{\sqrt{2}} e_{k_1}^{\mu_1} e_{k_2}^{\mu_2} V^{k_1 k_2 \bi} \\
        
        0 & 0 & e^{-1} e_{\bmu}^{\bi} 
    \end{bmatrix},\\
    E^{M}{}_{A} & = e^{-\phi}
    \begin{bmatrix}
        e^{\mu}_{j} & 0 & 0\\
        
        \frac{1}{\sqrt{2}}e^{\n}_{j} C_{\n \m_1 \m_2}  & e^{[j_1}_{\mu_1} e^{j_2]}_{\mu_2} & 0 \\
        
        - \frac{e}{2} e^{\bmu}_{j} U + \frac{e}{4} e_{j}^{\lambda} C_{\lambda \n \r} V^{\bmu \n \r} & \frac{e}{\sqrt{2}} V^{j_1 j_2 \bmu} & e e^{\bmu}_{\bj} 
    \end{bmatrix}.
\end{aligned}
\end{equation}
One lists some useful relations: $g = \det(g_{\m \n}) = (e^{-2\phi} e^{\frac23})^5\bg(y^{\mu}) = M^{-\frac{5}{27}} \bg(y^{\mu})$, $g_{\m\n} = M^{-\frac{1}{27}} \bg_{\m \n}$ $M^{\frac12} = E = det(E_{M}^{A}) = (e^{\phi} e^{-\frac13})^{27}$, $\mM^{\frac12} = \mE = det(\mE_{M}^{A}) = 1$, $\mM_{MN} = M^{-\frac{1}{27}} M_{MN}$.

Due to the rescaling the generalized vielbeins $E^{M}{}_{A}$ and $\mE^{M}{}_{A}$ transform with  different weights $\l[E^{M}{}_{A}] = \frac13$, $\l[\mE^{M}{}_{A}] = 0$ and hence one has
\begin{equation}\label{LieEtrunc}
    \mL_{\L} E^{M}{}_{A} = \L^K \dt_K E^{M}{}_{A} - E^{K}{}_{A} \dt_K \L^M  + 10 d^{MKR} d_{NLR} E^{N}{}_{A} \dt_K \L^L.
\end{equation}

Dropping all terms in the full Lagrangian of exceptional field theory which do not give contributions to the equations of motion of the generalised metric and $\det{g_{\m\n}}$ upon $A_\m{}^M=0,\, B_{\m\n M}=0$ one has
\begin{equation}\label{e6L}
     g^{-\frac12} \mL = \mc{R}[g_{\m \n}] + \mL_{sc}(\mM_{MN}, g_{\m \n})
\end{equation}
where $\hat{\mc{R}}[g_{\m \n}]$ is the usual Ricci scalar and the scalar potential is given by
\begin{equation}\label{e6Lsc}
\begin{aligned}
     \mL_{sc}(\mM_{MN}, g_{\m \n}) = & -\frac1{24} \mM^{MN} \dt_M \mM^{KL} \dt_N \mM_{KL}  -\frac1{2} \mM^{MN} \dt_M \mM^{KL} \dt_L \mM_{NK}\\
     & -\frac1{2} g^{-1} \dt_M g \dt_N \mM^{MN} - \frac1{4} \mM^{MN} g^{-2} \dt_M g \dt_N g - \frac1{4} \mM^{MN} \dt_M g^{\m\n} \dt_N g_{\m\n}.
\end{aligned}
\end{equation}
Upon the rescaling as above the Lagrangian can be written in the following simple form 
\begin{equation}\label{e6Ltrunc}
     \mL = \bg^{\frac12} M^{-\frac{1}{18}} (\mc{R}[\bg_{\m \n}] + \tilde{\mL}_{sc}(M_{MN})),
\end{equation}
where $\tilde{\mL}_{sc}(M_{MN})$ is the same as in the case of the non-linear realisation of E$_{6(6)}$ ExFT \cite{Berman:2011jh}. This due to the fact that the generalized metric obtained in the non-linear realisation has the same rescaling symmetry as our truncated metrics $M_{MN}$. 

Now one applies the same logic as in \cite{Bakhmatov:2020kul}, that is to notice that equations of motion of the rescaled generalised metric are those coming from the scalar potential $\tilde{\mL}_{sc}(M_{MN})$ plus 'cosmological term' coming from the curvature scalar of the external space $\mc{R}[\bg_{\m \n}]$
\begin{equation}
    \fr{\d \tilde{\mL}_{sc}(M_{MN})}{\d M_{MN}}  - \frac{1}{18}\tilde{\mL}_{sc}(M_{MN}) M^{MN} = \fr1{18} \mc{R}[\bg_{\m \n}]M^{MN}.
\end{equation}
In what follows we assume that these are satisfied for the undeformed background and search for conditions upon which the deformation does not spoil this.

In what follows it proves convenient to turn to the so-called flux formulation of the scalar sector of exceptional field theory and to rewrite the above Lagrangian in terms of generalised fluxes as in \cite{Musaev:2013rq}. Indeed, written completely in terms of components of fluxes $\mF_{AB}{}^C$, to be defined below, equations of motion will be guaranteed to hold after a deformation if the latter the flux components. Given the deformation is an E${}_{6(6)}$ transformation this would simply be a requirement for the flux components to transform covariantly. 

In \cite{Borsato:2020bqo} for generalised fluxes of Double Field Theory this requirement has been shown to be equivalent to the classical Yang-Baxter equation. The same idea is applicable here. Hence, one defines
\begin{equation}\label{LieE2}
    \mL_{E_{A}} E^{M}{}_{B} = E^{K}{}_{A} \dt_K E^{M}{}_{B} - E^{K}{}_{B} \dt_K E^{M}{}_{A}  + 10 d^{MKR} d_{NLR} E^{N}{}_{B} \dt_K E^{L}{}_{A} = \mF_{A,B}{}^{C} E^{M}{}_{C},
\end{equation}
with generalised flux components
\begin{equation}\label{Fstructure}
    \mF_{A,B}{}^{C} = 2 E_{M}{}^{C} E^{K}{}_{[A|} \dt_K E^{M}{}_{|B]} + 10 d^{MKR} d_{NLR} E_{M}{}^{C} E^{N}{}_{B} \dt_K E^{L}{}_{A}.
\end{equation}
Note that here one does not require the fluxes to be constant as it is done in generalised Scherk-Shwarz blackuction of \cite{Musaev:2013rq}. The latter generate the scalar potential of the maximal $D=5$ gauged supergravity, where preservation of supersymmetry requires the so-called linear constraint $\mF_{A,B}{}^{C} \in \bf{27} \, \oplus \, \bf{351}$. In components on has the trombone $\theta_{A} \in \bf{27}$ and  Z-flux $Z_{AB}{}^C \in \bf{351}$ \cite{deWit:2004nw,LeDiffon:2008sh}. 

The components $\mF_{A,B}{}^C$ defined in \eqref{LieE2} automatically satisfy this constraint and the corresponding components read
\begin{equation}\label{fluxes}
    \begin{aligned}
       -27 \theta_{A} & = 9 \partial_{M} E^{M}{}_{A} + E_{M}{}^{B} \partial_{A} E^{M}{}_{B}, \\
       Z_{A B}{}^{C} & = 10 d_{N L R} d^{M K R} E_{M}{}^{C} E^{N}{}_{(B|} E_{M}{}^{C} \dt_{K} E^{L}{}_{|A)}.
    \end{aligned}
\end{equation}
The second line here can be further simplified. Define first the symmetric invariant tensors $d_{ABC}$ and $d^{ABC}$ by explicitly listing the components  as in \eqref{Dtensors}
\begin{equation}\label{Dtensorsflat}
    \begin{aligned}
       d^{\ba a}{}_{b_1 b_2} & = \frac{1}{\sqrt{5}} \delta_{[b_1}{}^{\ba} \delta_{b_2]}{}^{a}, &&
       d_{a_1 a_2 a_3 a_4 a_5 a_6} &&& = \frac{1}{4\sqrt{5}} \epsilon_{a_1 a_2 a_3 a_4 a_5 a_6},\\
       d_{\ba a}{}^{b_1 b_2} & = \frac{1}{\sqrt{5}} \delta^{[b_1}{}_{\ba} \delta^{b_2]}{}_{a}, &&
       d^{a_1 a_2 a_3 a_4 a_5 a_6} &&& = \frac{1}{4\sqrt{5}} \epsilon^{a_1 a_2 a_3 a_4 a_5 a_6}.
    \end{aligned}
\end{equation}
Due to the non-vanishing weight of the  rescaled generalised vielbein  $E^{M}{}_{A} \in E_{6(6)} \times \mathbb{R}^{+}$ the invariant tensors in curved indices $d^{MNK}$ and $d_{MNK}$ are related to that with flat indices as
\begin{equation}\label{DtensorsEtrunc}
d_{M N P} E^{M}{}_{A} E^{N}{}_{B} E^{P}{}_{C} = e^{3 \phi} \, e \,d_{A B C}.
\end{equation}
Note, that for $\mc{E}_M{}^A$ one has the expected relation
\begin{equation}\label{DtensorsE}
d_{M N P} \mE^{M}{}_{A} \mE^{N}{}_{B} \mE^{P}{}_{C} = d_{A B C}.
\end{equation}
Using this we finally obtain the following expression for $Z$-flux
\begin{equation}\label{Zflux}
    \begin{aligned}
       Z_{A B}{}^{C} & = 5 d_{A B D} d^{M K R} E_{M}{}^{C} \Big(e^{3 \phi} e \partial_{K} E_{R}{}^{D} + E_{R}{}^{D} \partial_{K} (e^{3 \phi} e)\Big),
    \end{aligned}
\end{equation}
which will be used in what follows.

\section{Polyvector deformations}
\label{sec:defsflux}

\subsection{Transformation of fluxes}

In the non-linear realisation approach to construction of the generalised metric of exceptional field theory, one finds that generators of positive level $t^{m_1m_2 m_3}, t^{m_1\dots m_6}$  generate lower triangular generalised vielbein and the corresponding 3- and 6-forms give the standard field content of supergravity. Working with negative level generators $t_{m_1m_2 m_3}, t_{m_1\dots m_6}$ one replaces $p$-forms with $p$-vectors generalising the $\b$-frame of double field theory \cite{Riccioni:2009xr}, that proves convenient in describing non-geometric background (see e.g. \cite{Andriot:2013xca}). Since both such realisations give properly defined element of the coset $G/K$ with $G$ and $K$ being the global and local U-duality groups respectively, it is convenient to define a polyvector deformation as a linear  transformation given by the following E${}_{6(6)}$ element
\begin{equation}
    \label{eq:O_gen}
    O=\exp \Big[\fr{1}{3!}\W^{mnk}t_{mnk}\Big]\exp\Big[\fr{1}{6!}\W^{m_1\dots m_6}t_{m_1\dots m_6}\Big].
\end{equation}
 In components the deformation map $O \in \mathrm{E}_{6(6)}$ can be written as follows
\begin{equation}\label{ODefE6}
    O_{M}{}^{N} = 
    \begin{bmatrix}
        \delta_{m}{}^{n} & 0 & 0\\
        
        -\frac{1}{\sqrt{2}} \Omega^{n m_1 m_2}  & \delta^{m_1}{}_{[n_1} \delta^{m_2}{}_{n_2]} & 0 \\    
        \frac{e^{-1}}{2} \delta_{\bm}{}^{n} J + \frac{e^{-1}}{4}  \Omega^{n k l} W_{\bm k l} & - \frac{e^{-1}}{\sqrt{2}} W_{n_1 n_2 \bm} & \delta_{\bm}{}^{\bn} 
    \end{bmatrix},
\end{equation}
where we define $J = \frac{e}{6!} \epsilon_{m_1 \dots m_6} \Omega^{m_1 \dots m_6}$ and $W_{m_1 m_2 m_3} = \frac{e}{3!} \epsilon_{m_1 \dots m_6} \Omega^{m_4 m_5 m_6}$. Barblack small Latin indices are the same as unbarblack indices and have been introduced for technical convenience to distinguish between blocks of the generalised vielbein.

Hence, one defines deformation of the generalised vielbein and of its inverse as follows 
\begin{equation}
    \begin{aligned}
       E_{M}{}^{A} & \longrightarrow  O_{M}{}^{N} E_{N}{}^{A} = (\delta_{M}{}^{N} + \Omega_{M}{}^{N}) E_{N}{}^{A}, \\
     E^{M}{}_{A} & \longrightarrow  (O^{-1})^{M}{}_{N} E^{N}{}_{A} = (\delta^{M}{}_{N} + \tilde{\Omega}^{M}{}_{N}) E^{N}{} _{A},
    \end{aligned}
\end{equation}
where one singles out the unity matrix and  introduces covariant deformation tensors $\W_M{}^N$ and $\tilde{\Omega}^{M}{}_{N}$. These contain only 3- and 6-vector deformation parameters and one naturally decomposes deformation of the generalised fluxes in powers of $\W$'s. While for $\W_M{}^N$ the expression is self-evident, for $\tilde{\Omega}^{M}{}_{N}$ one obtains
\begin{equation}\label{OmegaDefE6inv}
        \tilde{\Omega}^{N}{}_{M} = 
    \begin{bmatrix}
        0 & \frac{1}{\sqrt{2}} \Omega^{n m_1 m_2} & - \frac{e^{-1}}{2} \delta_{\bm}{}^{n} J + \frac{e^{-1}}{4}  \Omega^{n k l} W_{\bm k l}\\
        
        0  & 0 & \frac{e^{-1}}{\sqrt{2}} W_{n_1 n_2 \bm} \\
        
        0 & 0 & 0
    \end{bmatrix}.
\end{equation}

One now considers transformation of generalised fluxes under such defined deformations. The simplest option one searches for here is when generalised fluxes (in flat indices) do not transform upon a condition on the 3- and 6-index $\r$-tensors. Given the flux formulation of the truncated exceptional field theory when all equations of motion of the scalar sector are written purely in terms of fluxes, that guarantees that the deformation does not spoil equations of motion. For double field theory this has been observed to be precisely the case when classical Yang-Baxter equation appears \cite{Borsato:2018idb}. Moreover, the condition on generalised fluxes in flat indices to transform as scalars appears as a natural condition for   fluxes in curved indices to transform covariantly. Indeed, consider
\begin{equation}
    \begin{aligned}
        \mF'{}_{MN}{}^K & = E'_M{}^AE'_N{}^B E'_C{}^K \mF'_{AB}{}^C\\
        &= O_M{}^P O_N{}^Q (O^{-1})^K{}_L \mF_{PQ}{}^L + \Delta \mF_{MN}{}^K,
    \end{aligned}
\end{equation}
where the additional non-covariant term $\Delta \mF_{MN}{}^K$ comes from a non-trivial transformation  of $\mF_{AB}{}^C$. Hence, one concludes that generalised fluxes $\mF_{AB}{}^C$ must naturally be scalars under such defined E${}_{6(6)}$ transformations. 

In general one might be interested in deriving conditions on general $\W^{mnk}$ and $\W^{m_1 \dots m_6}$ imposed by above constraints on transformations of fluxes. However here we restrict the narrative to only 3- and 6-Killing deformation, i.e. when 
\begin{equation}
    \begin{aligned}
         \W^{m_1 \dots m_3} &= \fr{1}{3!} \r^{i_1 \dots i_3} k_{i_1}{}^{m_1}     \cdots k_{i_3}{}^{m_3},\\
         \W^{m_1 \dots m_6} &=  \r^{i_1 \dots i_6} k_{i_1}{}^{m_1}     \cdots k_{i_6}{}^{m_6}.
    \end{aligned}
\end{equation}
 Here $\{k_i{}^m\}$ are Killing vectors of the initial undeformed background which satisfy an algebra defined by the usual expression
\begin{equation}
    k_{i_1}{}^m\dt_m k_{i_2}{}^n -k_{i_2}{}^m\dt_m k_{i_1}{}^n = f_{i_1i_2}{}^{i_3} k_{i_3}{}^n.
\end{equation}

Consider now transformation of the trombone flux $\q_A$, for which one finds
\begin{equation}
    \d \q_A = (...) {\rho}^{[i_1| i_2 i_3} {f}_{i_2 i_3}\,^{|i_4]} + (...) {\rho}^{[i_1 i_2 i_3 i_4| i_5 i_6} {f}_{i_5 i_6}\,^{|i_7]}.
\end{equation}
Here terms in brackets contain Killing vectors and various expressions in vielbein, and do not contain derivatives of Killing vectors. Hence, one may naturally impose the following sufficient conditions
\begin{equation}
    \label{eq:uni}
    \begin{aligned}
        {\rho}^{[i_1| i_2 i_3} {f}_{i_2 i_3}\,^{|i_4]}&=0, \\
        {\rho}^{[i_1 i_2 i_3 i_4| i_5 i_6} {f}_{i_5 i_6}\,^{|i_7]} &=0.
    \end{aligned}
\end{equation}
Recall, for deformation of  10d supergravity one has very similar conditions coming from the trombone flux  $r^{i_1 i_2} f_{i_1 i_2}{}^{i_3}=0$, where both lower indices of the structure constants get contracted with the $r$-matrix. This condition guarantees unimodularity of the dual algebra. Contracted with Killing vectors this gives the vector $I^m$ of generalised supergravity equations of motion. Hence this unimodularity constraint regulates whether one has a generalised or an ordinary supergravity background upon deformation, given CYBE is satisfied. 

One finds it natural to call the conditions \eqref{eq:uni} unimodularity constraints, which indeed ensure tracelessness of the dual structure constants. One may speculate here on generalised supergravity in 11-dimensions as well as on relaxing this condition, while keeping the flux invariant. Postponing this discussion to the Section \ref{sec:conclusions} we notice, that the above condition has been also found in the analysis of exceptional Drinfeld algebra and hence seems the most natural (see below).

While for invariance of the trombone flux the unimodularity condition is sufficient, transformation of the flux $Z_{AB}{}^C$ becomes more subtle. Here one finds it useful to consider the transformation order-by-order in the deformation tensor $\W_M{}^N$ or equivalently in $\r$-tensor, where $\r^{i_1\dots i_6}$ is understood as order 2. Hence, one obtains at order 1
\begin{equation}
    (\d Z_{AB}{}^C)_{\rm 1st} = (\dots) {\rho}^{i_1 i_2 i_3} {f}_{i_2 i_3}\,^{i_4},
\end{equation}
where as before terms in brackets contain only Killing vectors. Note the absence of an antisymmetrisation in $[i_1 i_4]$. From the explicit form of the transformation provided in the Appendix \ref{app:flux} one concludes, that no combination of $\r$-tensors found in terms of higher order can be used to cancel such first order terms. This forces to strengthen the unimodularity constraint found in the trombone flux and demand
\begin{equation}
     {\rho}^{i_1 i_2 i_3} {f}_{i_2 i_3}\,^{i_4}=0.
\end{equation}
Note, that the condition of precisely this form has been found in \cite{Malek:2020hpo}.

At order two one finds the following
\begin{equation}
    \begin{aligned}
        (\d Z_{AB}{}^C)_{\rm 2nd} = & \  (\dots) \rho^{[i_1 i_2 i_3 i_4| i_5 i_6} {f}_{i_5 i_6}\,^{|i_7]}\\
        &+ (\dots)_{[i_2 i_3 i_4 i_7]}  \Big({\rho}^{i_1 i_2 i_6} \rho^{i_3 i_4 i_5} f_{i_5 i_6}{}^{i_7} - \frac32 \rho^{i_2 i_3 i_4 i_5 i_6 i_7} {f}_{i_5 i_6}\,^{i_1}\Big)
    \end{aligned}
\end{equation}
where terms in the first line just reproduce the previously found unimodularity condition for $\r^{i_1 \dots i_6}$, and we left the indices $[i_2 i_3 i_4 i_7]$ on the terms in brackets in the second line to show explicitly the antisymmetry. One finds the terms in parentheses projected on the antisymmetric part in  $[i_2 i_3 i_4 i_7]$ is the condition, sufficient for the second order transformation to vanish. For further comparison with the results of \cite{Malek:2020hpo} let us rewrite this condition in more convenient form. Denoting the projection by subscript we have
\begin{equation}
    \begin{aligned}
    &\Big({\rho}^{i_1 i_2 i_6} \rho^{i_3 i_4 i_5} f_{i_5 i_6}{}^{i_7} - \frac32 \rho^{i_2 i_3 i_4 i_5 i_6 i_7} {f}_{i_5 i_6}\,^{i_1}\Big)_{[i_2i_3 i_4 i_7 ]}=\Big({\rho}^{i_1 i_2 i_6} \rho^{i_3 i_4 i_5} f_{i_5 i_6}{}^{i_7} -4 {f}_{i_5 i_6}\,^{[i_1}\rho^{i_2] i_3 i_4 i_5 i_6 i_7}\Big)_{[i_2i_3 i_4 i_7 ]}\\
    =&\fr43\Big({\rho}^{i_1 i_2 i_6} \rho^{i_3 i_4 i_5} f_{i_5 i_6}{}^{i_7}- {\rho}^{i_2 [i_1 |i_6|} \rho^{i_3 i_4 |i_5|} f_{i_5 i_6}{}^{i_7]} -3 {f}_{i_5 i_6}\,^{[i_1}\rho^{i_2] i_3 i_4 i_5 i_6 i_7}\Big)_{[i_2i_3 i_4 i_7 ]}.
    \end{aligned}
\end{equation}
Here in the first line we used the identity 
\begin{equation}\label{rhorelation3}
    5 {f}_{i_5 i_6}\,^{[i_1} {\rho}^{i_2 i_3 i_4 i_7] i_5 i_6} + 4 {f}_{i_5 i_6}\,^{[i_2|} {\rho}^{i_1 |i_3 i_4 i_7| i_5 i_6} - {f}_{i_5 i_6}\,^{i_1} {\rho}^{i_2 i_3 i_4 i_7 i_5 i_6} = 0
\end{equation}
and the unimodularity condition on the 6-tensor. In the second line we used the identity 
\begin{equation}\label{rhorelation4}
     0\equiv {f}_{i_5 i_6}\,^{[i_2} {\rho}^{i_1 i_3 |i_5|} \rho^{i_4 i_7] i_6}=
     \Big({f}_{i_5 i_6}\,^{i_2} {\rho}^{i_1 i_3 i_5} \rho^{i_4 i_7 i_6} - 4 {f}_{i_5 i_6}\,^{[i_1|} {\rho}^{i_2 |i_3| i_5} \rho^{|i_4 i_7] i_6}\Big)_{[i_2i_3 i_4 i_7 ]},
\end{equation}
where the most LHS is due to antisymemtry in $[i_5 i_6]$ and the RHS provides a convenient decomposition of the antisymmetrisation $[i_1i_2i_3i_4i_7]$. Hence, the sufficient condition that takes into account symmetries of the remaining terms in the transformation of the generalised flux reads
\begin{equation}
    \label{eq:genYB1}
        {\rho}^{i_1 [i_2 |i_6|} \rho^{i_3 i_4 |i_5|} f_{i_5 i_6}{}^{i_7]}- {\rho}^{i_2 [i_1 |i_6|} \rho^{i_3 i_4 |i_5|} f_{i_5 i_6}{}^{i_7]} -3 {f}_{i_5 i_6}\,^{[i_1}\rho^{i_2] i_3 i_4 i_5 i_6 i_7}=0.
\end{equation}
This has precisely the same for as the condition on the $\r$-tensors obtained in \cite{Malek:2020hpo} from analysis of exceptional Drinfeld algebra, however up to some additional identifications to be explicitly provided in Section \ref{sec:rel}.

Let us now turn to the analysis of terms in transformation of the flux $Z_{AB}{}^{C}$ of the last third order where one finds an additional constraint
\begin{equation}
       (\delta Z_{A B}{}^{C})_{\rm 3rd} = (\dots )_{ [i_3 i_4 i_5 i_6 i_7 i_{10}]}\Big(\rho^{i_1 i_2 [i_8} \rho^{i_3 i_4 i_9} \rho^{i_5 i_6 i_7]} f_{i_8 i_9}{}^{i_{10}}- 18 {\rho}^{i_1 i_2 [i_8} {\rho}^{i_3 i_4 i_5 i_6 i_7 i_9]} f_{i_8 i_9}{}^{i_{10}}\Big),
\end{equation}
where we drop all terms proportional to the unimodularity and the quadratic constraint \eqref{eq:genYB1}, and leave explicit indices  $[i_3 i_4 i_5 i_6 i_7 i_{10}]$ of terms in brackets. Hence, one might take the expression cubic in $\rho$-tensor as an additional constraint sufficient for the E${}_{6(6)}$ fluxes to be invariant
\begin{equation}
        \label{eq:genYB2}
       \rho^{i_1 i_2 [i_8} \rho^{i_3 i_4 i_9} \rho^{i_5 i_6 i_7]} f_{i_8 i_9}{}^{i_{10}}- 18 {\rho}^{i_1 i_2 [i_8} {\rho}^{i_3 i_4 i_5 i_6 i_7 i_9]} f_{i_8 i_9}{}^{i_{10}}=0,
\end{equation}
probably with additional antisymmetrisation in $[i_3 i_4 i_5 i_6 i_7 i_{10}]$. Notice that the above condition has $7$ antisymmetrised indices labelling generators of the symmetry algebra, defined by Killing vectors. This is the reason why such condition cannot appear in the approach of \cite{Malek:2020hpo} restricted to group manifolds, when $i,j=1,\dots,6$. One expects to find this additional constraint in the E${}_{7(7)}$ exceptional field theory.

Hence, we find the following conditions which are sufficent for generalised fluxes of the E${}_{6(6)}$ exceptional field theory to be invariant and hence for a deformation to be solution-generating
\begin{equation}
    \begin{aligned}
        {\rho}^{i_1 i_2 i_3} {f}_{i_2 i_3}\,^{i_4}&=0, \\
        {\rho}^{[i_1 i_2 i_3 i_4| i_5 i_6} {f}_{i_5 i_6}\,^{|i_7]} &=0,\\
         {\rho}^{i_1 [i_2 |i_6|} \rho^{i_3 i_4 |i_5|} f_{i_5 i_6}{}^{i_7]}- {\rho}^{i_2 [i_1 |i_6|} \rho^{i_3 i_4 |i_5|} f_{i_5 i_6}{}^{i_7]} -3 {f}_{i_5 i_6}\,^{[i_1}\rho^{i_2] i_3 i_4 i_5 i_6 i_7}&=0,\\
          {\rho}^{i_1 i_2 [i_8} {\rho}^{i_3 i_4 i_5 i_6 i_7 i_9]} f_{i_8 i_9}{}^{i_{10}}&=0.
    \end{aligned}
\end{equation}
First three conditions above are precisely the same as in the EDA approach of \cite{Malek:2020hpo}, while the last line is an additional condition, which can not be seen in the E${}_{6(6)}$ exceptional Drinfeld algebra description of group manifold backgrounds.

\subsection{Relation to the algebraic approach of EDA}
\label{sec:rel}

The supergravitational analysis above describes the deformation map $O_M{}^N$ encoding a 3- and 6-vector deformation of an 11-dimensional background. Using explicit relations between 11-dimensional fields and those of the E${}_{6(6)}$ exceptional field theory the corresponding generalised metric is defined, whose deformation is a linear transformation. Let us now provide explicit relation between the $r$-tensors and constraints one them appearing in deformation of supergravity backgrounds and  those appearing from constraint on consistency of deformations of the E${}_{6(6)}$ exceptional Drinfeld algebra \cite{Malek:2020hpo}.

One first notices, that parametrisation of the $\bf 27$  employed for the generalised metric here is different from that used to define the E${}_{6(6)}$ EDA in  \cite{Malek:2020hpo}. Indeed, here we use  $T_M=\{T_{m},T^{[m_1,m_2]},T_{\bm}\}$, while in \cite{Malek:2020hpo} one has
\begin{equation}
    T_M=\{T_{m},T^{[m_1,m_2]},T^{[m_1,m_2,m_3,m_4,m_5]}\}.
\end{equation}
In this form the origin of the last $\bf 6$ from windings of the M5-brane is more transparent. Hence, the relation is simply given by the invariant tensor of SL(6) 
\begin{equation}\label{transition}
    T_{\bm} = T^{[m_1 m_2 m_3 m_4 m_5]} \frac{1}{\sqrt{5!}} \epsilon_{\bm [m_1 m_2 m_3 m_4 m_5]}.
\end{equation}

Next, one compares the deformation map $O_M{}^N$ here and the matrix $C_M{}^N$ (denoted $C_A{}^B$ in the eq. (4.3) of \cite{Malek:2020hpo}. While the deformation map $O_M{}^N$ acts on the generalised vielbein and one requires such deformed generalised vielbein to encode a solution, the matrix $C_M{}^N$ acts on generators $\{T_M\}$ of exceptional Drinfeld algebra (EDA) and one requires the transformed generators to again form an EDA. One has 
\begin{equation}\label{OCcomparison1}
    \begin{aligned}
    O_{M}{}^{N}& = 
    \begin{bmatrix}
        \delta_{m}{}^{n} & 0 & 0\\
        
        -\frac{1}{\sqrt{2}} \Omega^{n m_1 m_2}  & \delta^{m_1}{}_{[n_1} \delta^{m_2}{}_{n_2]} & 0 \\
        
        \frac{e^{-1}}{2} \delta_{\bm}{}^{n} J + \frac{e^{-1}}{4}  \Omega^{n k l} W_{\bm k l} & - \frac{e^{-1}}{\sqrt{2}} W_{n_1 n_2 \bm} & \delta_{\bm}{}^{\bn} , 
    \end{bmatrix}, \\
    C_{M}{}^{N} & = 
    \begin{bmatrix}
        \delta_{n}{}^{m} & 0 & 0\\
        
        \frac{1}{\sqrt{2}} \brho^{n m_1 m_2}  & \delta^{a_1}{}_{[n_1} \delta^{m_2}{}_{n_2]} & 0 \\
        
        \frac{1}{\sqrt{5!}} \brho^{n m_1 ... m_5} + \frac{5}{\sqrt{5!}} \brho^{n [m_1 m_2} \brho^{m_3 m_4 m_5]} & \frac{20}{\sqrt{2!5!}} \delta^{[m_1}{}_{n_1} \delta^{m_2}{}_{n_2} \brho^{m_3 m_4 m_5]}  & \delta^{m_1}{}_{[n_1} ...\delta^{m_5}{}_{n_5]} 
    \end{bmatrix}.
    \end{aligned}
\end{equation}
where again $J = \frac{e}{6!} \epsilon_{m_1 \dots m_6} \Omega^{m_1 ... m_6}$ and $W_{m_1 m_2 m_3} = \frac{e}{3!} \epsilon_{m_1 \dots m_6} \Omega^{m_4 m_5 m_6}$. For group manifolds Killing vectors can be identified with the vielbein $e_i{}^m = k_i{}^m$ and one has
\begin{equation}
    \begin{aligned}
     \W^{m_1 \dots m_3} &= \fr{1}{3!} \r^{i_1 i_2 i_3} k_{i_1}{}^{m_1}k_{i_2}{}^{m_2}k_{i_3}{}^{m_3} = \fr16 \r^{m_1 m_2 m_3},\\
     \W^{m_1 \dots m_6} &= \r^{i_1 \dots i_6} k_{i_1}{}^{m_1}\dots k_{i_6}{}^{m_6} = \r^{m_1 \dots  m_6},
    \end{aligned}
\end{equation}
i.e. the $\r$-tensors now acquire space (curved) indices. Now using $[\mu_1...\mu_7]=0$ one  rewrites
\begin{equation}\label{dJrewrite}
    \delta_{\bm}{}^{n} J = \delta_{\bm}{}^{n} \frac{e}{6!} \epsilon_{n_1 \dots n_6} \rho^{n_1 \dots n_6} = \frac{e}{5!} \epsilon_{\bm n_1 \dots n_5} \rho^{n n_1 \dots n_5}.
\end{equation}
Hence, the deformation map for group manifolds becomes
\begin{equation}\label{OCcomparison2}
    O_{M}{}^{N} = 
    \begin{bmatrix}
        \delta_{m}{}^{n} & 0 & 0\\
        
        -\frac{1}{6\sqrt{2}} \rho^{n m_1 m_2}  & \delta^{m_1}{}_{[n_1} \delta^{m_2}{}_{n_2]} & 0 \\
        
        \frac{1}{5!2} \epsilon_{\bm n_1 ... n_5} \rho^{n n_1...n_5} + \frac{1}{4\cdot6^3}  \rho^{n k l} \epsilon_{\bm k l n_1 n_2 n_3} \rho^{n_1 n_2 n_3}& - \frac{1}{36 \sqrt{2}} \epsilon_{m_1 m_2 \bm n_1 n_2 n_3} \rho^{n_1 n_2 n_3} & \delta_{\bm}{}^{\bn} 
    \end{bmatrix}.
\end{equation}
Comparing this to the matrix $C_M{}^N$ above and using \eqref{transition} one finds
\begin{equation}
\label{connectionofrhos}
    \rho^{m_1 m_2 m_3} = - 6 \brho^{m_1 m_2 m_3},\qquad \rho^{m_1 m_2 m_3 m_4 m_5 m_6} = 2 \brho^{m_1 m_2 m_3 m_4 m_5 m_6}.
\end{equation}

Now having at hands equivalence between transformations encoded by the matrices $O$ and $C$ one may investigate relations between the constraints on the $\r$-tensors. Consistency constraints for the generators $T_M$ transformed by the matrix $C_M{}^N$ to form an EDA derived in \cite{Malek:2020hpo}  read 
\begin{equation}\label{EQUATIONSMalec}
    \begin{aligned}
          {\brho}^{m_1 m_2 m_3} {f}_{m_1 m_2}\,^{m_4}& = 0, && \text{unimodularity}_3,\\
          {\brho}^{m_1 m_2 [m_3 \dots m_6} {f}_{m_1 m_2}\,^{m_7]} &= 0, && \text{unimodularity}_6,\\
         3 f_{k_1 k_2}{}^{[n_1} {\brho}^{n_2 n_3] k_1}   {\brho}^{m_1 m_2 k_2} - 5 f_{k_1 k_2}{}^{[m_1} {\brho}^{m_2] [n_1 n_2} {\brho}^{n_3 k_1 k_2]} - f_{k_1 k_2}{}^{[m_1} {\brho}^{m_2] n_1 n_2 n_3 k_1 k_2}&=0 , && \mbox{``gen. CYBE''}.
    \end{aligned}
\end{equation}
While the unimodularity conditions in the first two lines are precisely the same as the ones following from flux invariance, the third line requires more work. Let us expand the antisymmetrisation in $[ n_1 n_2 n_3 k_1 k_2]$ of the third line above and reorganise it the following more symmetric form
\begin{equation}
    6 f_{k_1 k_2}{}^{[n_1} {\brho}^{n_2 n_3 | k_1|} {\brho}^{m_1] m_2 k_2} - 6 f_{k_1 k_2}{}^{[n_1} {\brho}^{n_2 n_3 | k_1|} {\brho}^{m_2] m_1 k_2} = f_{k_1 k_2}{}^{[m_1} {\brho}^{m_2] n_1 n_2 n_3 k_1 k_2} .
\end{equation}
Upon \eqref{connectionofrhos} this becomes precisely \eqref{eq:genYB1}.

\subsection{The short SL(5) story}

Let us now briefly look at 3-vector deformations in the formalism of the SL(5) exceptional field theory and show that the only condition appearing both in the algebraic and ExFT approaches is the unimodularity constraint. 

Here we are working in the same conventions as that of \cite{Bakhmatov:2019dow} and for the sake of brevity we will avoid lengthy description of the SL(5)-covariant exceptional field theory. For details of the  construction of the SL(5) Exceptional Field Theory see \cite{Berman:2010is,Musaev:2015ces}. Important however is to mention the generalised Lie derivative of the generalised vielbein
\begin{equation}
    \begin{aligned}
    \mL_\L E^{M}{}_{A } &= 
    \frac{1}{2}\, {\L}^{KL}  {\partial}_{KL}{{E}^{M}\,_{A}}\,  + {\partial}_{KL}{{\L}^{M K}}\,  {E}^{L}\,_{A}+ \frac{1}{4}\, {\partial}_{KL}{{\L}^{KL}}\,  {E}^{M}\,_{A},
    \end{aligned}
\end{equation}
whose explicit form tells that we are working in the truncated theory. Here $M, N, K,\dots =1,\dots, 5$ label coordinate indices while $A, B, C, \dots =1\dots 5$ label flat indices. As before small Latin indices $m, n, k, \dots$ and $a, b, c,\dots $ label directions of the ``internal'' space, that is four-dimensional in this case. 

Explicitly the generalised metric and the deformation map are given by
\begin{equation}
\label{O}
    E_M{}^{A} = e^{\fr{\f}{2}}
        \begin{bmatrix}
			e^{-1/4} e_m{}^{a} && e^{1/4} v^a \\ \\
			0 &&  e^{1/4}
		\end{bmatrix}, \quad	
	     O_M{}^N = \begin{bmatrix}
      \d_m{}^n && 0 \\
      \\
      \fr1{3!}\e_{mpqr}\W^{pqr} && 1
    \end{bmatrix},
\end{equation}
where $v^m=\fr1{3!}\e^{mnkl}C_{nkl}$ and as before $    \W^{m_1m_2m_3}= \fr1{3!}\r^{i_1i_2i_3}k_{i_1}{}^{m_1}k_{i_2}{}^{m_2}k_{i_3}{}^{m_3}$.
Generalised flux is defined as structure constants of the corresponding Leibniz algebra as follows
\begin{equation}
     \mL_{E_{AB}}E^{M}{}_{C}=\mF_{AB,C}{}^{D}E^{M}{}_{D},
\end{equation}
where $E_{AB}{}^{MN}=2E_{[A}{}^ME_{B]}{}^N$. In \cite{Berman:2012uy} it has been shown explicitly that such defined flux contains only the $\bf 10$, $\bf 15$ and $\bf \overline{40}$ of SL(5), that in tensor notations are encoded by the trombone $\q_{[AB]}$ and by fluxes $Y_{(AB)}$ and $Z^{AB,C}$. Explicitly one defines the irblackucible flux components as $Y_{(BC)} = \mF_{A(B,C)}{}^{A}$, $\theta_{[BC]} = \mF_{A[B,C]}{}^{A}$, $\mF_{[ABC]}{}^{D} = Z_{ABC}{}^{D} + \theta_{[AB} \delta_{C]}^{D}$, that  gives
\begin{equation}
    \begin{aligned}
    \q_{AB} &=E^{-1}E^{MN}{}_{AB}\dt_{MN}E - E^M{}_{[A} \dt_{MN}E^N{}_{B]}, \\
    Y_{AB}&=E^M{}_{(A} \dt_{MN}E^N{}_{B)},\\
    Z_{ABC}{}^{D}& = E^{M}_{[A} E^{N}_{B|} E_{K}^{D} \partial_{M N}{E^{K}_{|C]}} + \frac13 \Big(2 E^{M}_{[A|} \partial_{M N}{E^{N}_{|B|}} + E^{M}_{[A} E^{N}_{B|} E^{-1} \partial_{M N}{E}\Big) \delta_{|C]}^{D}.
    \end{aligned}
\end{equation}
Here we denote $E=\det E^A{}_M$. Following the same procedure as for the E${}_{6(6)}$ case one finds that transformation of all components of the generalised flux at all orders in $\r^{i_1i_2i_3}$ can be written as
\begin{equation}
    \d \mF_{AB,C}{}^D= (\dots)_{[i_1i_4]} \r^{i_1 i_2 i_3}f_{i_2 i_3}{}^{i_4}.
\end{equation}
Note, that the unimodularity condition here is more relaxed, than that of the E$_{6(6)}$ case. Hence the unimodularity constraint is indeed sufficient for the flux to be invariant and for the  deformation to generate a solution.

At the algebraic side one considers the SL(5) exceptional Drinfeld algebra developed in \cite{Sakatani:2019zrs,Malek:2019xrf} and deforms generators as follows
\begin{equation}
    \begin{aligned}
        T_m &\to T_m,\\
        T^{mn} & \to T^{mn}+ \rho^{mnk} T_k.
    \end{aligned}
\end{equation}
Requiring the deformed generators to also form an SL(5) exceptional Drinfeld algebra one derives the following constraints (see \cite{Sakatani:2019zrs} for more details)
\begin{equation}
    \label{eq:constr_SL5}
    \begin{aligned}
       \r^{m_1m_2m_3} f_{m_2 m_3}{}^{m_4}&=0,\\
       4 f_{k_1 k_2}{}^{[m_1}\r^{m_2] k_1 [n_1}\r^{n_2 n_3] k_2} + 3 f_{k_1 k_2}{}^{[n_1}\r^{n_2 n_3]k_1}\r^{m_1 m_2 k_2}&=0.
    \end{aligned}
\end{equation}
The first line is simply the unimodularity constraint whose appearance was expected. The second line is a quadratic constraint, which is however equivalent to the first line. Indeed, since the indices $m,n,k=1,\dots 4$ it is natural to define
\begin{equation}
    \r_m=\e_{mnkl}\rho^{nkl}.
\end{equation}
Substituting this into the second line of \eqref{eq:constr_SL5} and contracting the indices $m_1 m_2$ and $n_1 n_2$ with epsilon-tensor one rewrites the constraint simply as
\begin{equation}
    \r_{m_1} \r_{[n_1} f_{n_2 n_3]}{}^{m_1}=0,
\end{equation}
which vanishes identically upon the unimodularity constraint. Note that while one is able to define such $\r_m$ only on group manifold, we see that the unimodularity constraint is still sufficient for a tri-vector SL(5) deformation to generate solutions.

One notices however, that the non-abelian deformation provided in \cite{Bakhmatov:2020kul} are all non-unimodular, while still  provide solutions. This nicely illustrates the fact that such obtained conditions are only sufficient, not necessary. At the same time this observation again raises the question of searching for non-trivial tri-vector deformations, now satisfying generalised classical Yang-Baxter equation.

\section{Conclusions and discussions}
\label{sec:conclusions}

In this work we investigate 3- and 6-Killing deformations of general backgrounds of 11-dimensional supergravity admitting at least three Killing vectors, not necessarily commuting. The formalism of exceptional field theory provides more natural degrees of freedom for that than the conventional formulation of supergravity. In these terms tri-vector deformations appear to be encoded by an E${}_{d(d)}$ element generated by a 3- and 6-vector. Since the field content of ExFT is given by tensor of GL(11-d) taking values in irreps of the U-duality group, these transform linearly. Schematically that means
\begin{equation}
    \label{eq:gen_def}
    \Phi_{\mu_1\dots \m_p}{}^{M_1 \dots M_N} \to O_{N_1}{}^{M_1} \cdots O_{N_N}{}^{M_N}\Phi_{\mu_1\dots \m_p}{}^{N_1 \dots N_N}.
\end{equation}
Here the indices $\m=1\dots 11-d$, the indices $M,N$ label certain irrep of E${}_{d(d)}$ and the matrix $O$ is defined as in \eqref{eq:O_gen}, and we restrict the polyvectors to be proportional to Killing vectors of the initial background
\begin{equation}
    \begin{aligned}
         \W^{m_1 \dots m_3} &= \fr{1}{3!} \r^{i_1 \dots i_3} k_{i_1}{}^{m_1}     \cdots k_{i_3}{}^{m_3},\\
         \W^{m_1 \dots m_6} &=  \r^{i_1 \dots i_6} k_{i_1}{}^{m_1}     \cdots k_{i_6}{}^{m_6}.
    \end{aligned}
\end{equation}
Working for concreteness in the E${}_{6(6)}$ exceptional field theory we consider its truncation to the scalar sector as in \cite{Bakhmatov:2019dow}. This leaves us with only the generalised metric parametrising the coset E${}_{6(6)}\times \RR^+/{\rm Usp}(8)$. Explicitly the corresponding generalises vielbein is given in \eqref{uptrE6R}. While the generalised vielbein transforms under the deformations linearly, transformation of the supergravity fields entering its definition are very non-obvious. The covariant formalism allows us to refrain from attempting to provide  explicit formulae for deformations of supergravity fields as it was done \cite{Bakhmatov:2019dow}. Such analysis however would be necessary for one to generate explicit examples of deformations within the E${}_{6(6)}$ theory. 

Truncated exceptional field theory can be completely written in terms of generalised fluxes, as it has been explicitly shown in \cite{Musaev:2013rq} for the E${}_{6(6)}$ symmetry group. The corresponding equations of motion for the generalised vielbein can as well be written  in terms of fluxes with a single factor of the inverse vielbein $E_A{}^M$. Hence, the sufficient condition for a deformation to map a solution of such equations of motion to a solution is the condition for generalised fluxes to transform covariantly. Hence fluxes with local USp(8) indices $\mF_{AB}{}^C$ must be invariant. We investigate transformation of generalised fluxes under such defined transformations and learn that one may ensure invariance if a condition on the constant tensors $\r^{i_1i_2i_3} $ and $\r^{i_1 \dots i_6}$. The latter are a generalisation of the classical $r$-matrix and the condition is believed to be a generalisation of the classical Yang-Baxter equation. We show that the condition sufficient for the deformation to be a solution-generating transformation is precisely that derived via $\r$-deformation of the E${}_{6(6)}$ exceptional Drinfeld algebra in \cite{Malek:2019xrf}.

The same analysis of generalised fluxes of the truncated SL(5) theory shows that the sufficient condition is simply the unimodularity constraint on $\r^{i_1i_2i_3}$, i.e.
\begin{equation}
    \r^{i_1i_2i_3}f_{i_2i_3}{}^{i_4}=0.
\end{equation}
We notice that the quadratic constraint on $\r^{i_1i_2i_3}$ derived in \cite{Sakatani:2019zrs} is satisfied identically upon the unimodularity constraint, that provides consistency between supergravity and EDA picture.

It is worth  here to return back to the result of \cite{Bakhmatov:2019dow} where explicit examples of tri-vector deformations of AdS${}_4\times \SS^7$ based on the SL(5) approach have been presented. One checks that all non-abelian examples there are non-unimodular, that naively seems to contradict the above result. However, since the LHS of the unimodularity condition gets contracted with subtle expressions in vielbeins, 3-forms and Killing vectors, the condition is far from being necessary. One concludes, that the results of \cite{Bakhmatov:2019dow} provide examples of tri-Killing deformations which are not (generalised) Yang-Baxter in the above sense. To our knowledge the literature does not contain examples of such non-Yang-Baxter \emph{bi}-vector deformations relevant for two-dimensional sigma-models. It would be interesting to search for examples of such deformations in more space-time dimensions, where one has non-trivial CYBE in addition to the unimodularity constraint.

Given the explicit check of invariance of (internal) generalised fluxes of the E${}_{6(6)}$ exceptional field theory where at the end of the day subtle and long terms  boil down to relatively simple expressions containing the algebraic equation, one might expect the same to happen for all fields of any exceptional field theory. Since as we discussed above, the examples of tri-vector deformations of \cite{Bakhmatov:2019dow} stand slightly apart from the standard narrative of Yang-Baxter deformations, it is still an interesting problem to find examples of deformation, which do satisfy generalised Yang-Baxter equation. Another interesting direction of further research is to investigate tri-vector deformations of 10-dimensional supergravity and in particular of two-dimensional sigma-models, which raises intriguing questions of integrability of such deformations. This is work in progress and results will be reported in further publications.

\section*{Acknowledgements}

This work has been supported by Russian Science Foundation grant RSCF-20-72-10144. Authors acknowledge discussions with I. Bakhmatov, E. Malek, N. S. Deger that motivated this project.

\appendix

\section{Notations and conventions}\label{NotesRefs}

In this paper we use the following conventions for indices
\begin{equation}
    \begin{aligned}
       &\hat{\mu}, \hat{\nu}, = 1 \dots 11&& \mbox{eleven directions, curved}; \\
       &\hat{\alpha}, \hat{\beta}, = 1 \dots 11&& \mbox{eleven directions, flat}; \\       
       &\mu, \nu, \rho, \ldots = 1 \dots 5&& \mbox{external  directions of ExFT, curved}; \\
       &\bar{\mu}, \bar{\nu}, \bar{\rho}, \ldots = 1 \dots 5&& \mbox{external   directions of ExFT, flat}; \\  
       &k, l, m, n,  \ldots = 1,\dots,6 && \mbox{internal six directions, curved}; \\
       &\bar{k}, \bar{l}, \bm, \bn,  \ldots = 1,\dots,6 && \mbox{internal six directions, curved}; \\       
       & a, b, c, d, \ldots = 1,\dots,6 && \mbox{internal six directions, flat}; \\       
       & M, N, K, L, \ldots = 1,\dots, 27 && \mbox{fundamental ExFT indices, curved};  \\
       & A, B, C, D, \ldots = 1,\dots, 27 && \mbox{fundamental ExFT indices, flat};\\
       & i, j = 1,\dots, N && \mbox{indices labelling Killing vectors}; \\
    \end{aligned}
\end{equation}

\section{Derivation of flux transformations }
\label{app:flux}

Tedious part of calculations involved with explicit substitution of the deformed vielbein into flux expressions, decomposition and preliminary simplification of the result have been done using the computer algebra program Cadabra v.1.4 \cite{Cadabra,Peeters:2007wn} (note that v.2 is also available \cite{Peeters:2018dyg}). The corresponding Cadabra files can be found here \cite{git:fluxE6}

One starts with the trombone flux whose transformation becomes
\begin{equation}\label{defTrombone}
    \begin{aligned}
       \delta \theta_{A} = & - 3\, \frac{1}{\sqrt{2}} {E}_{m n A} {k}_{i_1}\,^{m} {\color{black}{\rho}^{[i_1| i_2 i_3} {f}_{i_2 i_3}\,^{|i_4]}} {k}_{i_4}\,^{n} + \frac{1}{96}\, {Ee}^{k}\,_{A} {\epsilon}_{k l m n p q} {k}_{i_1}\,^{l} {k}_{i_2}\,^{m} {k}_{i_3}\,^{n} {k}_{i_4}\,^{p} {\rho}^{i_1 i_2 i_3} {\color{black}{\rho}^{[i_4| i_5 i_6} {f}_{i_5 i_6}\,^{|i_7]}} {k}_{i_7}\,^{q}\\
       & - \frac{3}{16}\,  {Ee}^{k}\,_{A} {\epsilon}_{k l m n p q} {k}_{i_1}\,^{l} {k}_{i_2}\,^{m} {k}_{i_3}\,^{n} {k}_{i_4}\,^{p} {\color{black}{\rho}^{[i_1 i_2 i_3 i_4| i_5 i_6} {f}_{i_5 i_6}\,^{|i_7]}} {k}_{i_7}\,^{q} \\
       & + \frac{1}{96}\, {Ee}^{k}\,_{A} {\epsilon}_{k l m n p q} {k}_{i_1}\,^{l} {k}_{i_2}\,^{m} {k}_{i_3}\,^{n} {k}_{i_4}\,^{p} {\rho}^{i_1 i_2 i_3} {\color{black}{\rho}^{i_5 i_6 [i_7|} {f}_{i_5 i_6}\,^{|i_4]}} {k}_{i_7}\,^{q} \sim \\
       & \sim (...) \underbrace{{\color{black}{\rho}^{[i_1| i_2 i_3} {f}_{i_2 i_3}\,^{|i_4]}}}_{\text{unimodularity}_3} + (...) \underbrace{{\color{black}{\rho}^{[i_1 i_2 i_3 i_4| i_5 i_6} {f}_{i_5 i_6}\,^{|i_7]}}}_{\text{unimodularity}_6}.
    \end{aligned}
\end{equation}
One observes that terms of both first and second order in the $\r$-tensors are proportional to the unimodularity constraint. Note, that although here te constraint comes with antisymmetrisation of the upper indices, further analysis of Z-flux will require more strict version of the constraint without the antisymmetrisation
\begin{equation}
    {\rho}^{i_2 i_3 i_4} {f}_{i_2 i_3}\,^{i_1}=0.
\end{equation}

Transformation of the Z-flux is convenient to analyse order-by-order in the $\r$-tensor assyming that $\r^{i_1i_2i_3}$ is of order one while $\r^{i_1\dots i_6}$ is of order two. Hence, at first order one has
\begin{equation}\label{defZ1}
    \begin{aligned}
       (\delta Z_{A B}{}^{C})_{\rm 1st}   = & \frac{5e}{6\sqrt{10}}\,  {d}_{A B D\, }\Big( {E}_{
      \bm}\,^{C} {E}_{n}\,^{D\, } {k}_{i_1}\,^{\bm} {\color{black}{\rho}^{i_2 i_3 [i_4|} {f}_{i_2 i_3}\,^{|i_1]}} {k}_{i_4}\,^{n}    + \,   {E}_{\bm}\,^{[D\, } {E}_{n}\,^{C]} {k}_{i_1}\,^{\bm} {\color{black}{\rho}^{i_2 i_3 i_4} {f}_{i_2 i_3}\,^{i_1}} {k}_{i_4}\,^{n}\\
       & + \frac{1}{4}\,  {E}^{k l C} {E}^{m n D\, } {\epsilon}_{k l m n p q} {k}_{i_1}\,^{p}  {\color{black}{\rho}^{i_2 i_3 [i_1} f_{i_2 i_3}{}^{i_4]}} {k}_{i_4}\,^{q} \Big),
    \end{aligned}
\end{equation}
which again vanishes upon the unimodularity constraint.

Next, at order two one writes
\begin{equation}\label{defZ2}
    \begin{aligned}
        (\d Z_{AB}{}^C)_{\rm 2nd}=
       &  \frac{e}{24\sqrt{5}}  {d}_{A B D\, } {\epsilon}_{l m n p q r}\Big(\frac{1}{12}\,  {E}_{k}\,^{(D|\, } {E}^{l m |C)} \,{i_1}\,^{k}\,_{i_2}\,^{n}\,_{i_3}\,^{p}\,_{i_4}\,^{q} {\rho}^{i_1 [i_2 i_3|} {\color{black}{\rho}^{i_5 i_6 |i_7} {f}_{i_5 i_6}\,^{i_4]} } \\
       &  + \frac{1}{18}\,  {E}_{k}\,^{D\, } {E}^{l m C}  K_{i_1}\,^{k}\,_{i_2}\,^{n}\,_{i_3}\,^{p} \,_{i_4}\,^{q} {\rho}^{i_2 i_3 i_4} {\color{black}{\rho}^{i_5 i_6 [i_1} {f}_{i_5 i_6}\,^{i_7]}} \\
       &  + \frac{1}{36}\,   {E}_{k}\,^{[D|\, } {E}^{l m |C]} K_{i_1}\,^{k}\,_{i_5}\,^{p} \,_{i_6}\,^{q}\,{i_4}\,^{n}  {\rho}^{i_5 i_6 i_7} {\color{black}{\rho}^{i_1 i_2 i_3} {f}_{i_2 i_3}\,^{i_4}}  \\
       &  + \frac{1}{6}\,    {E}_{k}\,^{[D|\, } {E}^{l m |C]}  K_{i_1}\,^{k}\,_{i_2}\,^{n}\,_{i_3}\,^{p} \,_{i_4}\,^{q}  {\color{black}{\rho}^{i_1 [i_2| i_6} {\rho}^{|i_3 i_4| i_5} f_{i_5 i_6}{}^{|i_7]}}  \\
       & - \frac{1}{4}\,  {E}_{k}\,^{D\, } {E}^{l m C}  K_{i_1}\,^{k}\,_{i_2}\,^{n}\,_{i_3}\,^{p}\,_{i_4}\,^{q} {\color{black}{\rho}^{i_2 i_3 i_4 i_5 i_6 i_7} {f}_{i_5 i_6}\,^{i_1}}  \\
       & - \,  {E}_{k}\,^{(D|\, } {E}^{l m |C)} K_{i_1}\,^{k}\,_{i_2}\,^{n}\,_{i_3}\,^{p}\,_{i_4}\,^{q} {\color{black}{\rho}^{i_1 [i_2 i_3 i_4| i_5 i_6} {f}_{i_5 i_6}\,^{|i_7]}}   \Big)  \,{i_7}\,^{r},
    \end{aligned}
\end{equation}
where for clarity of expressions we define
\begin{equation}
    K_{i_1}{}_{ \dots}^{m_1}{}^{\dots}_{i_p}{}^{m_p}= k_{i_1}{}^{m_1}\cdots k_{i_p}{}^{m_p}.
\end{equation}
Given the unimodularity constraint the first three lines vanish and the above expression becomes
\begin{equation}\label{defZ2oder}
    \begin{aligned}
       (\d Z_{AB}{}^C)_{\rm 2nd}=
       &  \frac{e}{24\sqrt{5}}  {d}_{A B D\, } {\epsilon}_{l m n p q r}\Big( \frac{1}{6}\,    {E}_{k}\,^{[D|\, } {E}^{l m |C]} K_{i_1}\,^{k}\,_{i_2}\,^{n}\,_{i_3}\,^{p}\,_{i_4}\,^{q}  {\color{black}{\rho}^{i_1 [i_2| i_6} {\rho}^{|i_3 i_4| i_5} f_{i_5 i_6}{}^{|i_7]}}  \\
       & - \frac{1}{4}\,  {E}_{k}\,^{D\, } {E}^{l m C} K_{i_1}\,^{k}\,_{i_2}\,^{n}\,_{i_3}\,^{p}\,_{i_4}\,^{q} {\color{black}{\rho}^{i_2 i_3 i_4 i_5 i_6 i_7} {f}_{i_5 i_6}\,^{i_1}}  \\
       & - \,  {E}_{k}\,^{(D|\, } {E}^{l m |C)} K_{i_1}\,^{k}\,_{i_2}\,^{n}\,_{i_3}\,^{p}\,_{i_4}\,^{q} {\color{black}{\rho}^{i_1 [i_2 i_3 i_4| i_5 i_6} {f}_{i_5 i_6}\,^{|i_7]}}   \Big) \,_{i_7}\,^{r}.
    \end{aligned}
\end{equation}
Using the identity
\begin{equation}\label{rhorelation}
    {\rho}^{i_1 [i_2 i_3 i_4| i_5 i_6} {f}_{i_5 i_6}\,^{|i_7]} = \frac54 {\rho}^{[i_1 i_2 i_3 i_4| i_5 i_6} {f}_{i_5 i_6}\,^{|i_7]} - \frac14 {\rho}^{i_2 i_3 i_4 i_7 i_5 i_6} {f}_{i_5 i_6}\,^{i_1},
\end{equation}
and performing a series of algebraic manipulations the final expression can be massaged to \begin{equation}\label{defZ2odercontinue}
    \begin{aligned}
        &(\d Z_{AB}{}^C)_{\rm 2nd}= \\
        &- \frac{5\sqrt{5} e}{96}{d}_{A B D\, } {E}_{k}\,^{(D|\, } {E}^{l m |C)} {\epsilon}_{l m n p q r}K_{i_1}\,^{k}\,_{i_2}\,^{n}\,_{i_3}\,^{p}\,_{i_4}\,^{q}\,_{i_7}{}^{r} {\color{black} {\rho}^{[i_1 i_2 i_3 i_4| i_5 i_6} {f}_{i_5 i_6}\,^{|i_7]}} \\
       & + \frac{\sqrt{5}e}{3 \cdot 48} {d}_{A B D\, } {E}_{k}\,^{[D|\, } {E}^{l m |C]} {\epsilon}_{l m n p q r} K_{i_1}\,^{k}\,_{i_2}\,^{n}\,_{i_3}\,^{p}\,_{i_4}\,^{q}\,_{i_7}\,^{r}  \Big({\color{black}{\rho}^{i_1 [i_2| i_6} {\rho}^{|i_3 i_4| i_5} f_{i_5 i_6}{}^{|i_7]} - \frac32 {\rho}^{i_2 i_3 i_4 i_5 i_6 i_7} {f}_{i_5 i_6}\,^{i_1}})\Big) .
    \end{aligned}
\end{equation}
The yellow expression in the first line is the unimodularity condition for the tensor $\r^{i_1 \dots i_6}$, while the black terms in the second line compose the generalised classical Yang-Baxter equation. The latter is equivalent to the constraint obtained in \cite{Malek:2020hpo}.

Finally, terms of third order in the $\rho$-tensors read
\begin{equation}\label{defZ}
    \begin{aligned}
        &(\d Z_{AB}{}^C)_{\rm 3d}= \fr{e}{96\sqrt{10}} d_{ABD}\times\\
       &  \times\Big(\frac{5}{54}\,   {E}_{k}\,^{C} {E}_{l}\,^{D\, } {\epsilon}_{m n p q r t}K_{i_1}\,^{k}\,_{i_{10}}\,^{l}\,_{i_2}\,^{m}\,_{i_3}\,^{n}\,_{i_4}\,^{p}\,_{i_5}\,^{q}\,_{i_6}\,^{r} {\rho}^{i_1 i_2 i_3} {\rho}^{i_4 i_5 i_6} {\color{black}{\rho}^{i_7 i_8 i_9} {f}_{i_7 i_8}\,^{i_{10}}}\,_{i_9}\,^{t} \\
       &  +  \frac{5}{54}\,   {E}_{k}\,^{(C} {E}_{l}\,^{D)\, } {\epsilon}_{m n p q r t}K_{i_1}\,^{k}\,_{i_{10}}\,^{l}\,_{i_2}\,^{m}\,_{i_3}\,^{n}\,_{i_4}\,^{p}\,_{i_5}\,^{q}\,_{i_6}\,^{r}\,_{i_9}\,^{t} {\rho}^{i_1 i_2 i_3} {\rho}^{i_5 i_6 i_9} {\color{black}{\rho}^{i_{10} i_7 i_8} {f}_{i_7 i_8}\,^{i_4}} \\
       & + \frac{1}{18}\,   {E}_{k}\,^{C} {E}_{l}\,^{D\, } {\epsilon}_{m n p q r t}K_{i_1}\,^{k}\,_{i_{10}}\,^{l}\,_{i_2}\,^{m}\,_{i_3}\,^{n}\,_{i_4}\,^{p}\,_{i_5}\,^{q}\,_{i_6}\,^{r}\,_{i_9}\,^{t} {\rho}^{i_2 i_3 i_4 i_5 i_6 i_9} {\color{black}{\rho}^{i_7 i_8 [i_{10}} {f}_{i_7 i_8}\,^{i_1]}}        \\
       & - 5\sqrt{2}  {E}_{k}\,^{D\, } {E}^{l m C} {\epsilon}_{l m n p q r}K_{i_1}\,^{k}\,_{i_2}\,^{n}\,_{i_3}\,^{p}\,_{i_4}\,^{q}\,_{i_7}\,^{r} {\color{black}{\rho}^{i_2 i_3 i_4 i_5 i_6 i_7} {f}_{i_5 i_6}\,^{i_1}} \\
       & - 4\sqrt{2},   {E}_{k}\,^{(D|\, } {E}^{l m |C)} {\epsilon}_{l m n p q r}K_{i_1}\,^{k}\,_{i_2}\,^{n}\,_{i_3}\,^{p}\,_{i_4}\,^{q} {\color{black}{\rho}^{i_1 [i_2 i_3 i_4| i_5 i_6} {f}_{i_5 i_6}\,^{|i_7]}}\,_{i_7}\,^{r}\\
       & + \frac{5}{27}\,    {E}_{k}\,^{C} {E}_{l}\,^{D\, } {\epsilon}_{m n p q r t}K_{i_1}\,^{k}\,_{i_2}\,^{l}\,_{i_3}\,^{m}\,_{i_4}\,^{n}\,_{i_5}\,^{p}\,_{i_6}\,^{q}\,_{i_7}\,^{r}\,_{i_{10}}\,^{t}\,  {\color{black}{\rho}^{i_1 [i_3| i_8} {\rho}^{i_2 |i_4| i_9} {\rho}^{|i_5 i_6 i_7} f_{i_8 i_9e}{}^{i_{10}]}}  \\
       & - \frac{5}{54}\,    {E}_{k}\,^{C} {E}_{l}\,^{D\, } {\epsilon}_{m n p q r t}K_{i_1}\,^{k}\,_{i_2}\,^{l}\,_{i_3}\,^{m}\,_{i_4}\,^{n}\,_{i_5}\,^{p}\,_{i_6}\,^{q}\,_{i_7}\,^{r}\,_{i_{10}}\,^{t} \,  {\color{black}{\rho}^{i_1 [i_3| i_8} {\rho}^{|i_{10} i_4| i_9} {\rho}^{|i_5 i_6 i_7]} f_{i_8 i_9e}{}^{i_2}} \\
       & + \frac{5}{18}\,    {E}_{k}\,^{C} {E}_{l}\,^{D\, } {\epsilon}_{m n p q r t}K_{i_1}\,^{k}\,_{i_2}\,^{l}\,_{i_3}\,^{m}\,_{i_4}\,^{n}\,_{i_5}\,^{p}\,_{i_6}\,^{q}\,_{i_7}\,^{r}\,_{i_{10}}\,^{t} \, {\color{black}{\rho}^{i_1 [i_3| i_8}{\rho}^{|i_7 i_4| i_9}  {\color{black}{\rho}^{|i_5 i_6| i_2}} f_{i_8 i_9e}{}^{|i_{10}]}} \\
        & + \frac{1}{3}\,    {E}_{k}\,^{C} {E}_{l}\,^{D\, } {\epsilon}_{m n p q r t}K_{i_1}\,^{k}\,_{i_2}\,^{l}\,_{i_3}\,^{m}\,_{i_4}\,^{n}\,_{i_5}\,^{p}\,_{i_6}\,^{q}\,_{i_7}\,^{r}\,_{i_{10}}\,^{t}  \,  {\color{black}{\rho}^{i_1 i_2 i_9} {\rho}^{[i_3 i_4 i_5 i_6 i_7| i_8} f_{i_8 i_9e}{}^{|i_{10}]}}  \Big).
    \end{aligned}
\end{equation}
The first five lines vanish because of the unimodularity constraint while the rest can be nicely packaged as follows 
\begin{equation}\label{defZ3oder}
    \begin{aligned}
       (\delta Z_{A B}{}^{C})_{3rd} & = \frac{\sqrt{5}e}{5184\sqrt{2}}  {d}_{A B D\, } {E}_{k}\,^{C} {E}_{l}\,^{D\, } {\epsilon}_{m n p q r t} K_{i_1}\,^{k}\,_{i_2}\,^{l}\,_{i_3}\,^{m}\,_{i_4}\,^{n}\,_{i_5}\,^{p}\,_{i_6}\,^{q}\,_{i_7}\,^{r}\,_{i_{10}}\,^{t}\, \bigg( 2 {{\rho}^{i_1 [i_3| i_8} {\rho}^{i_2 |i_4| i_9} {\rho}^{|i_5 i_6 i_7} f_{i_8 i_9}{}^{i_{10}]}}\\
       &  - {{\rho}^{i_1 [i_3| i_8} {\rho}^{|i_{10} i_4| i_9} {\rho}^{|i_5 i_6 i_7]} f_{i_8 i_9}{}^{i_2}} + 3 {{\rho}^{i_1 [i_3| i_8}} {{\rho}^{|i_7 i_4| i_9}  {\rho}^{|i_5 i_6| i_2} f_{i_8 i_9}{}^{|i_{10}]}} + 18 {{\rho}^{i_1 i_2 i_9} {\rho}^{[i_3 i_4 i_5 i_6 i_7| i_8} f_{i_8 i_9}{}^{|i_{10}]}} \bigg) .
    \end{aligned}
\end{equation}
Now, using the identity
\begin{equation}\label{rhorelation2}
    \begin{aligned}
    {\rho}^{i_1 [i_3| i_8} \rho^{|i_2 i_4| i_9} \rho^{|i_5 i_6 i_7} {f}_{i_8 i_9}\,^{i_{10}]} & = \frac27 {\rho}^{i_1 [i_3| i_8} \rho^{i_2 |i_4| i_9} \rho^{|i_5 i_6 i_7} {f}_{i_8 i_9}\,^{i_{10}]} - \frac17 {\rho}^{i_1 i_2 i_8} \rho^{[i_3 i_4| i_9} \rho^{|i_5 i_6 i_7} {f}_{i_8 i_9}\,^{i_{10}]}\\ 
    & - \frac37 {\rho}^{i_1 [i_3| i_8} \rho^{|i_7 i_4| i_9} \rho^{|i_5 i_6| i_2} {f}_{i_8 i_9}\,^{|i_{10}]} - \frac17 {\rho}^{i_1 [i_3| i_8} \rho^{|i_{10} i_4| i_9} \rho^{|i_5 i_6 i_7]} {f}_{i_8 i_9}\,^{i_2} ,
    \end{aligned}
\end{equation}
for the first two terms in (\ref{defZ3oder}), we obtain
\begin{equation}\label{defZ3odercontinue}
    \begin{aligned}
       &(\delta Z_{A B}{}^{C})_{3rd}=\\
       &\frac{e\sqrt{5}}{5184\sqrt{2}}  {d}_{A B D\, } {E}_{k}\,^{C} {E}_{l}\,^{D\, } {\epsilon}_{m n p q r t} K_{i_1}\,^{k}\,_{i_2}\,^{l}\,_{i_3}\,^{m}\,_{i_4}\,^{n}\,_{i_5}\,^{p}\,_{i_6}\,^{q}\,_{i_7}\,^{r}\,_{i_{10}}\,^{t}\, \bigg( {7 {{\rho}^{i_1 [i_3| i_8} \rho^{|i_2 i_4| i_9} \rho^{|i_5 i_6 i_7} {f}_{i_8 i_9}\,^{i_{10}]}}}\\
       & + {6 {{\rho}^{i_1 [i_3| i_8}} {{\rho}^{|i_7 i_4| i_9}  {{\rho}^{|i_5 i_6| i_2}} f_{i_8 i_9}{}^{|i_{10}]}}}  + {{\rho}^{i_1 i_2 i_8}} {\rho^{[i_3 i_4| i_9} \rho^{|i_5 i_6 i_7} {f}_{i_8 i_9}\,^{i_{10}]}} - 18 {{\rho}^{i_1 i_2 i_8}} { {\rho}^{[i_3 i_4 i_5 i_6 i_7| i_9} f_{i_8 i_9}{}^{|i_{10}]}} \bigg) .
    \end{aligned}
\end{equation}
Due to the high amount of index symmetries provided by contraction with the epsilon tensor, one is able to show that the first two terms are zero while the third and the fourth can be rewritten in a nice form. 

Indeed, start with the first term
\begin{equation}\label{1ofZ3}
\begin{aligned}
       &7K_{i_1}{}^{k}{}_{ i_2}{}^{l}{}_{ i_3}{}^{[m}{}_{ i_4} {}^{n}{}_{ i_5} {}^{p}{}_{  i_6} {}^{q}{}_{  i_7} {}^{r}{}_{  i_{10}} {}^{t]} {{\rho}^{i_1 [i_3| i_8} \rho^{|i_2 i_4| i_9} \rho^{|i_5 i_6 i_7} {f}_{i_8 i_9}\,^{i_{10}]}} \\
       &= 7 K_{i_1}{}^{k}{}_{i_2}{}^{[l}{}_{i_3}{}^{m}{}_{i_4}{}^{n}{}_{i_5 }{}^{p}{}_{i_6}{}^{q}{}_{i_7}{}^{r}{}_{i_{10}}{}^{t} {\rho}^{i_1 i_3 i_8} \rho^{i_2 i_4 i_9} \rho^{i_5 i_6 i_7} {f}_{i_8 i_9}\,^{i_{10}} = 0,
      \end{aligned}
\end{equation}
due to the antisymmetrisation in seven indices $[l m n p q r t]=0$ each running from 1 to 6.

For the second term we will use the generalized Yang-Baxter equation obtained from (\ref{defZ2odercontinue}), and the the following identity
\begin{equation}\label{rhorelation5}
    \begin{aligned}
    0 & = \rho^{[i_5 i_6| i_2} \rho^{|i_3 i_7 i_4| i_9 i_8 |i_{10}} f_{i_8 i_9}{}^{i_1]}  = \frac17 \rho^{[i_5 i_6| i_2} \rho^{|i_3 i_7 i_4| i_9 i_8 |i_{10}]} f_{i_8 i_9}{}^{i_1} \\
    & - \frac47 \rho^{[i_5 i_6| i_2} \rho^{i_1 |i_7 i_4| i_9 i_8 |i_{10}} f_{i_8 i_9}{}^{i_3]}.
    \end{aligned}
\end{equation}
Taken together with the identity $\rho^{[i_3 i_6| i_8} \rho^{|i_7 i_4] i_9} f_{i_8 i_9}{}^{i_{10}} = 0$ this implies
\begin{equation}\label{rhorelation22}
     K_{i_1 i_2 [i_3 i_4 i_5 i_6 i_7 i_{10}]}^{k l m n p q r t} \rho^{i_5 i_6 i_2} \rho^{i_3 [i_1| i_8} \rho^{|i_7 i_4| i_9} f_{i_8 i_9}{}^{|i_{10}]} = 0.
\end{equation}
Hence, we write for the second term in \eqref{defZ3oderfinal}
\begin{equation}\label{2ofZ3}
\begin{aligned}
    &6 K_{i_1}{}^{k}{}_{i_2}{}^{l}{}_{i_3}{}^{[m}{}_{i_4}{}^{n}{}_{i_5}{}^{p}{}_{i_6}{}^{q}{}_{i_7}{}^{r}{}_{i_{10}}{}^{t} {{\rho}^{[i_5 i_6| i_2}} {{\rho}^{i_1 |i_3| i_8}} {{\rho}^{|i_7 i_4| i_9}   f_{i_8 i_9}{}^{|i_{10}]}} \\
    &\stackrel{\text{gen.CYBE}}{=} 9K_{i_1}{}^{k}{}_{i_2}{}^{l}{}_{i_3}{}^{[m}{}_{i_4}{}^{n}{}_{i_5}{}^{p}{}_{i_6}{}^{q}{}_{i_7}{}^{r}{}_{i_{10}}{}^{t]}  \rho^{[i_5 i_6| i_2} \rho^{|i_3 i_7 i_4| i_9 i_8 |i_{10}]} f_{i_8 i_9}{}^{i_1} \\
    &= 36 K_{i_1}{}^{k}{}_{i_2}{}^{l}{}_{i_3}{}^{[m}{}_{i_4}{}^{n}{}_{i_5}{}^{p}{}_{i_6}{}^{q}{}_{i_7}{}^{r}{}_{i_{10}}{}^{t]} \rho^{[i_5 i_6| i_2} \rho^{i_1 |i_7 i_4| i_9 i_8 |i_{10}} f_{i_8 i_9}{}^{i_3]} \\
    &= 36 K_{i_1}{}^{k}{}_{i_2}{}^{l}{}_{i_3}{}^{[m}{}_{i_4}{}^{n}{}_{i_5}{}^{p}{}_{i_6}{}^{q}{}_{i_7}{}^{r}{}_{i_{10}}{}^{t]}  \rho^{i_5 i_6 i_2} \rho^{i_1 i_7 i_4 i_9 i_8 i_{10}} f_{i_8 i_9}{}^{i_3}  \\
   &\stackrel{\text{genCYBE}}{=} 36K_{i_1}{}^{k}{}_{i_2}{}^{l}{}_{[i_3}{}^{[m}{}_{i_4}{}^{n}{}_{i_5}{}^{p}{}_{i_6}{}^{q}{}_{i_7}{}^{r}{}_{i_{10}]}{}^{t} \rho^{i_5 i_6 i_2} \rho^{i_3 [i_1| i_8} \rho^{|i_7 i_4| i_9} f_{i_8 i_9}{}^{|i_{10}]} = 0.
\end{aligned}\end{equation}
Where the last identity is due to \eqref{rhorelation22}.

Next we consider the third term in \eqref{defZ3oderfinal}. Again using the unimodularity condition and the identity  $\rho^{[i_3 i_6| i_8} \rho^{|i_7 i_4] i_9} f_{i_8 i_9}{}^{i_{10}} = 0$ one derives the following useful relation
\begin{equation}\label{rhorelation8}
    \frac27K_{i_1}{}^{k}{}_{i_2}{}^{l}{}_{i_3}{}^{[m}{}_{i_4}{}^{n}{}_{i_5}{}^{p}{}_{i_6}{}^{q}{}_{i_7}{}^{r}{}_{i_{10}}{}^{t]}\rho^{i_1 i_2 i_8} \rho^{i_3 i_4 i_9} \rho^{i_5 i_6 i_7} f_{i_8 i_9}{}^{i_{10}} =K_{i_1}{}^{k}{}_{i_2}{}^{l}{}_{i_3}{}^{[m}{}_{i_4}{}^{n}{}_{i_5}{}^{p}{}_{i_6}{}^{q}{}_{i_7}{}^{r}{}_{i_{10}}{}^{t]} \rho^{i_1 i_2 [i_8} \rho^{i_3 i_4 i_9} \rho^{i_5 i_6 i_7]} f_{i_8 i_9}{}^{i_{10}}.
\end{equation}
This allows to rewrite the third term as
\begin{equation}\label{3ofZ3}
       K_{i_1}{}^{k}{}_{i_2}{}^{l}{}_{i_3}{}^{[m}{}_{i_4}{}^{n}{}_{i_5}{}^{p}{}_{i_6}{}^{q}{}_{i_7}{}^{r}{}_{i_{10}}{}^{t]} {{\rho}^{i_1 i_2 i_8}} {\rho^{[i_3 i_4| i_9} \rho^{|i_5 i_6 i_7} {f}_{i_8 i_9}\,^{i_{10}]}} {=} \frac72 K_{i_1}{}^{k}{}_{i_2}{}^{l}{}_{i_3}{}^{[m}{}_{i_4}{}^{n}{}_{i_5}{}^{p}{}_{i_6}{}^{q}{}_{i_7}{}^{r}{}_{i_{10}}{}^{t]}  {\rho^{i_1 i_2 [i_8} \rho^{i_3 i_4 i_9} \rho^{i_5 i_6 i_7]} f_{i_8 i_9}{}^{i_{10}}}.
\end{equation}

Finally, for the last term in \eqref{defZ3oderfinal} we will need the identity
\begin{equation}\label{rhorelation10}
    \frac27 K_{i_1}{}^{k}{}_{i_2}{}^{l}{}_{i_3}{}^{[m}{}_{i_4}{}^{n}{}_{i_5}{}^{p}{}_{i_6}{}^{q}{}_{i_7}{}^{r}{}_{i_{10}}{}^{t]} \rho^{i_1 i_2 i_8} \rho^{i_3 i_4 i_5 i_6 i_7 i_9} f_{i_8 i_9}{}^{i_{10}} = K_{i_1}{}^{k}{}_{i_2}{}^{l}{}_{i_3}{}^{[m}{}_{i_4}{}^{n}{}_{i_5}{}^{p}{}_{i_6}{}^{q}{}_{i_7}{}^{r}{}_{i_{10}}{}^{t]} \rho^{i_1 i_2 [i_8} \rho^{i_3 i_4 i_5 i_6 i_7 i_9]} f_{i_8 i_9}{}^{i_{10}},
\end{equation}
which is true upon the unimodularity constraint on $\r^{i_1 \dots i_6}$. This gives for the fourth term 
\begin{equation}\label{4ofZ3}
\begin{aligned}
       &- 18 K_{i_1}{}^{k}{}_{i_2}{}^{l}{}_{i_3}{}^{[m}{}_{i_4}{}^{n}{}_{i_5}{}^{p}{}_{i_6}{}^{q}{}_{i_7}{}^{r}{}_{i_{10}}{}^{t]}{{\rho}^{i_1 i_2 i_8}} { {\rho}^{[i_3 i_4 i_5 i_6 i_7| i_9} f_{i_8 i_9}{}^{|i_{10}]}} \\
       &= - K_{i_1}{}^{k}{}_{i_2}{}^{l}{}_{[i_3}{}^{m}{}_{i_4}{}^{n}{}_{i_5}{}^{p}{}_{i_6}{}^{q}{}_{i_7}{}^{r}{}_{i_{10}]}{}^{t} 63 {{\rho}^{i_1 i_2 [i_8} {\rho}^{i_3 i_4 i_5 i_6 i_7 i_9]} f_{i_8 i_9}{}^{i_{10}}}.
\end{aligned}
\end{equation}

Altogether, the third order terms in the transformation of the Z-flux can be written as
\begin{multline}\label{defZ3oderfinal}
       \delta (Z_{A B}{}^{C})_{3rd} = \frac{7\sqrt{5} e}{10368\sqrt{2}}\, e6  {d}_{A B D\, } {E}_{k}\,^{C} {E}_{l}\,^{D\, } {\epsilon}_{m n p q r t} K_{i_1}{}^{k}{}_{[i_2}{}^{l}{}_{i_3}{}^{m}{}_{i_4}{}^{n}{}_{i_5}{}^{p}{}_{i_6}{}^{q}{}_{i_7}{}^{r}{}_{i_{10}]}{}^{t}\,\times \\
       \times\bigg( {\rho^{i_1 i_2 [i_8} \rho^{i_3 i_4 i_9} \rho^{i_5 i_6 i_7]} f_{i_8 i_9}{}^{i_{10}}} - 18 {{\rho}^{i_1 i_2 [i_8} {\rho}^{i_3 i_4 i_5 i_6 i_7 i_9]} f_{i_8 i_9}{}^{i_{10}}} \bigg).
\end{multline}
We see, that terms organise into the expression in brackets, that is antisymmetric in seven indices $i$ labelling Killing vectors. For group manifolds, which have $i=1,\dots,6$, the thrid order terms would identically vanish. However, for general 6-dimensional manifolds with at least 7 Killing vectors we find an additional constraint. It is important to note, that the terms in brackets cannot in general be blackuced to the generalised Yang-Baxter equation \eqref{eq:genYB1}. Indeed, in the second term due to the antisymmetrisation the indices $i_8,i_9$  stay either both on $\r^{i_1\dots i_6} $ or separate between $\r^{i_1\dots i_3}$ and $\r^{i_1\dots i_6}$. The latter case is evidently beyond something which can be massaged into \eqref{eq:genYB1}.

\bibliographystyle{unsrturl}
\addcontentsline{toc}{section}{References}
\bibliography{biblio.bib}

\end{document}